\documentclass[%
 reprint,
superscriptaddress,
amsmath,amssymb,
aps,
pra,
]{revtex4-2}
\usepackage{caption}
\usepackage{subcaption}
\usepackage{graphicx}
\usepackage{xcolor}
\usepackage{graphicx}
\usepackage{dcolumn}
\usepackage{bm}
\usepackage{upgreek}
\usepackage{lineno}

\begin{document}

\preprint{APS/123-QED}
\title{Optical spatiotemporal Fourier synthesis: Tutorial}

\author{Murat Yessenov}
\email{Corresponding authors: yessenov@ucf.edu}
\affiliation{Harvard John A. Paulson School of Engineering and Applied Sciences, Harvard University, Cambridge, MA, USA}
\affiliation{CREOL, The College of Optics \& Photonics, University of Central Florida, Orlando, Florida 32816, USA}

\author{Ayman F. Abouraddy}
\email{raddy@creol.ucf.edu}
\affiliation{CREOL, The College of Optics \& Photonics, University of Central Florida, Orlando, Florida 32816, USA}


\begin{abstract}
Fourier synthesis is one of the foundations of physical optics. Spatial Fourier optics is a basis for understanding optical imaging, microscopy, and holography. In conventional Fourier optics, the complex spatial field distribution in the Fourier plane constitutes the spatial spectrum of the field to be realized in physical space. Analogously, in temporal Fourier optics the complex temporal spectrum can be manipulated for ultrafast pulse-shaping. We present here a tutorial on the emerging field of spatiotemporal Fourier optics whereby the spatial and temporal spectra are manipulated jointly to produce spatiotemporally structured optical fields that display unique propagation characteristics. In this tutorial, we focus on a subset of the overall class of non-separable spatiotemporally structured fields; namely, cylindrically symmetric fields in which each radial spatial frequency is associated with a single wavelength. This subset of fields comprises propagation-invariant wave packets that travel rigidly in linear media at a tunable group velocity, and includes space-time wave packets and other closely related structured fields. We describe a spatiotemporal Fourier synthesis system capable of preparing arbitrary optical fields belonging to this subclass.
\end{abstract}

\maketitle

\section{Introduction}

Fourier optics is one of the foundational pillars of physical optics that provides a well-established set of concepts and techniques for the spatial manipulation of optical fields \cite{GoodmanBook05,SalehBook07}. This topic was researched heavily in the 1960's and 1970's, and yielded a host of advances in optical imaging \cite{Ersoy07Book,Easton10Book,Scott15Book,Khare23Book}, holography \cite{Winthrop65PL,Stroke65APL,Winthrop66PL,Mustafi23Photonics}, signal and image processing \cite{ONeill56IRE,VanderLugt64IEEE,Lee81Book}, optical computing \cite{Preston72Book,Arsenault89Book,Caulfield98Computer,Ambs10AOT,McMahon23NRP}, among a variety of other applications \cite{Stark81Book,Shamir99Book}. Fourier optics was further enriched by the introduction of the fractional Fourier transform, which led to new applications that make use of the representation of the optical signal in spaces that are intermediate between physical space and Fourier space \cite{Mendlovic93JOSAA,Ozaktas93JOSAA,Lohmann93JOSAA,Ozaktas95JOSAA,Ozaktas01Book}. Fourier optics is perhaps no longer a subject for cutting-edge research at the level of its fundamentals. Nevertheless, it remains a foundation on which much of our understanding of optical imaging, microscopy, and holography is built. More recently, the availability of high-performance spatial light modulators (SLMs) \cite{Neff90ProcIEEE,Maurer11LPR}, micro-mirror arrays \cite{Jang04OE,Brennesholtz08Book,Hellman19OE}, diffractive optics \cite{Buralli91AO,Gil03JVSTB,Banerji19Optica}, metasurfaces \cite{Yu11Science}, multi-plane light converters \cite{Morizur10JOSAA,Fontaine19NC,Lib22PRA,MartinezBecerril24arxiv}, and photonic lanterns \cite{BlandHawthorn09OE,LeonSaval10OE,Davenport21OME} have enabled convenient processing of complex optical fields directly in the spatial domain without resort to the Fourier domain. This topic is typically denoted `structured light' \cite{Forbes21NP}, and has also spurred research towards directly producing such light fields from a laser cavity \cite{Ngcobo13NC,Naidoo16NP,Maguid18ACSP,Piccardo22JO,Ginis23NC,Forbes24NPR,Yaraghi25RFC}. In other words, spatial structuring of an optical field can now be readily performed in either the physical domain or in the Fourier domain. 

On an independent front, the pursuit of ever-shorter laser pulses led to the development of picosecond and femtosecond pulsed lasers \cite{Diels06Book,Weiner09Book}. Because electronic devices cannot perform at such short time scales, temporally modulating ultrashort pulses requires operating in the spectral domain (wavelength space), leading to a \textit{temporal} analog of conventional \textit{spatial} Fourier optics \cite{Weiner88JOSAB,Weiner00RSI,Weiner2011OC}. This process has been further enriched by the study of so-called space-time analogies or space-time duality \cite{Tournois64CRAS,Akhmanov69SPJETP,Kolner89OL,Kolner94IEEEJQE,Salem13AOP}, in which the analogs of configurations that are well-known in spatial Fourier optics are appropriated for temporal Fourier optics. In contrast to spatial Fourier optics, where the optical field can be modulated either directly in the spatial domain or in the Fourier domain, no current technological pathways exist to directly modulate ultrashort pulses in the time domain.

Over the past decade, broad efforts have been directed to hybrid combinations of spatial and temporal Fourier optics, which yield new forms of spatiotemporally structured light, thereby leading to a host of exciting research ideas and potential applications \cite{Yessenov22AOP}. These newly emerging ideas have roots extending back to Brittingham's proposal for a propagation-invariant pulsed beam called a focused-wave mode (FWM) \cite{Brittingham83JAP}, which requires introducing an intricate spatiotemporal structure into the optical field. By propagation invariance we refer to an optical field whose spatiotemporal profile travels rigidly without diffraction or dispersion (thus retaining the same shape and scale) at a constant group velocity. Despite wide-ranging theoretical efforts \cite{Besieris89JMP,Shaarawi89JAP,Heyman89IEEE,Ziolkowski91PRA,Ziolkowski1991ProcIEEE,Ziolkowski1992IEEETAP,Shaarawi95JMP,Shaarawi00JPA,Christodoulides04OE,Zamboni09PRA}, only a few successful experimental efforts emerged because of the continuing difficulty of inculcating precise spatiotemporal structure into a pulsed laser beam. The first demonstration of a rationally synthesized spatiotemporally structured optical field was the optical X-wave in 1997 \cite{Saari97PRL}, in which an ultrashort pulse is directed to a conventional Bessel-beam generator \cite{Reivelt03arxiv,Grunwald2000OL,Alexeev02PRL,Grunwald2003PRA,Bowlan09OL,Bonaretti09OE}, whereas FWMs have yet to be conclusively synthesized in the optical regime \cite{Reivelt00JOSAA,Reivelt02PRE,Yessenov19PRA}. Nevertheless, X-waves have not found applications in optics \cite{Turunen10PO}, in contrast to their usefulness in ultrasonics \cite{Lu92IEEEa,Lu92IEEEb,FigueroaBook14}, because an extremely large temporal bandwidth is required to observe the characteristic X-shaped structure \cite{Yessenov22AOP}. Although nonlinear optical interactions yield related field structures through enforced phase-matching conditions \cite{DiTrapani03PRL,Faccio06PRL,Faccio07OE,Faccio2007Book}, these nonlinear X-waves require high-energy pulses, and the phase-matching conditions are not sufficiently strict for precise spatiotemporal spectral structuring (as recently confirmed in \cite{Stefanska23ACSP}).

The past few years have witnessed the emergence of a new frontier that may be called `spatiotemporal Fourier optics.' This refers to the joint modulation of the spatial and temporal degrees-of-freedom of the optical field. Rather than relying on nonlinear optics, spatiotemporal Fourier optics provides alternative linear approaches to produce spatiotemporally structured light. For example, spatiotemporal spectral modulation \cite{Kondakci17NP} can reproduce nonlinear X-waves \cite{DiTrapani03PRL,Faccio2007Book}, which are in fact a subset of recently developed `space-time wave packets' (STWPs) \cite{Yessenov22AOP}, in which each temporal frequency $\omega$ (or wavelength) underpinning the temporal pulse profile is associated with a single spatial frequency (or transverse wave number) underpinning the transverse spatial beam profile \cite{Kondakci16OE,Kondakci17NP,Kondakci18OE}. Similarly, spatiotemporal optical vortices (STOVs) that carry transverse orbital angular momentum (OAM), after their initial discovery in ultrashort laser pulse collapse in material media \cite{Jhajj16PRX}, were subsequently synthesized via linear spatiotemporal spectral modulation \cite{Hancock19Optica,Chong20NP,Hancock21Optica}. STWPs and STOVs are two examples among many other of spatiotemporally structured light fields, which have witnessed tremendous growth over the past few years (see Discussion).

Spatiotemporal Fourier optics is a nascent field with much to be explored. We present here a basic framework for investigating this area, and focus on STWPs as a representative class of spatiotemporally structured fields. In the most general case, each frequency $\omega$ is associated with a prescribed spatial spectrum; alternatively, each spatial frequency is associated with a different temporal spectrum. In physical space, each point in time is associated with a different spatial beam profile; or each point in the spatial beam profile has a different temporal pulse profile. A general spatiotemporal Fourier synthesizer is still lacking, but progress has been made toward synthesizing such generalized spatiotemporally structured fields \cite{Yessenov22AOP,Piccardo22JO,Shen23JO}. We will focus here on the restricted case in which each temporal frequency is associated with a single spatial frequency, which includes STWPs and other propagation-invariant pulsed beams. This will help lay the foundation for future comprehensive studies of the more generalized scenarios of spaiotemporal Fourier optics that are currently being pursued.

This Tutorial is organized as follows. We first briefly review conventional Fourier optics for monochromatic light to establish the formalism and nomenclature, and apply it to so-called `diffraction-free' beams. From there, we introduce the general concept of spatiotemporal Fourier optics and then specialize to the subclass of azimuthally symmetric pulsed optical fields in which each temporal frequency is associated with a single spatial frequency. Several examples are examined: pulsed Bessel beams, X-waves, and STWPs. We then explore the experimental synthesis of such novel spatiotemporally structured fields, in which the field propagates invariantly in free space at a tunable group velocity. This is the first instantiation of a spatiotemporal Fourier synthesizer that is amenable to producing this subclass of fields. In the Discussion, we widen our perspective to encompass the full space of spatiotemporally structured fields, enumerate the various phenomena that ensue from such field structuring, and discuss the possibility of generalizing the specialized spatiotemporal Fourier synthesizer described here to realize such fields.

\section{Fourier optics with monochromatic light}

\subsection{Basic concept of Fourier optics}

\begin{figure}[t!]
    \centering
    \includegraphics[width=8.6cm]{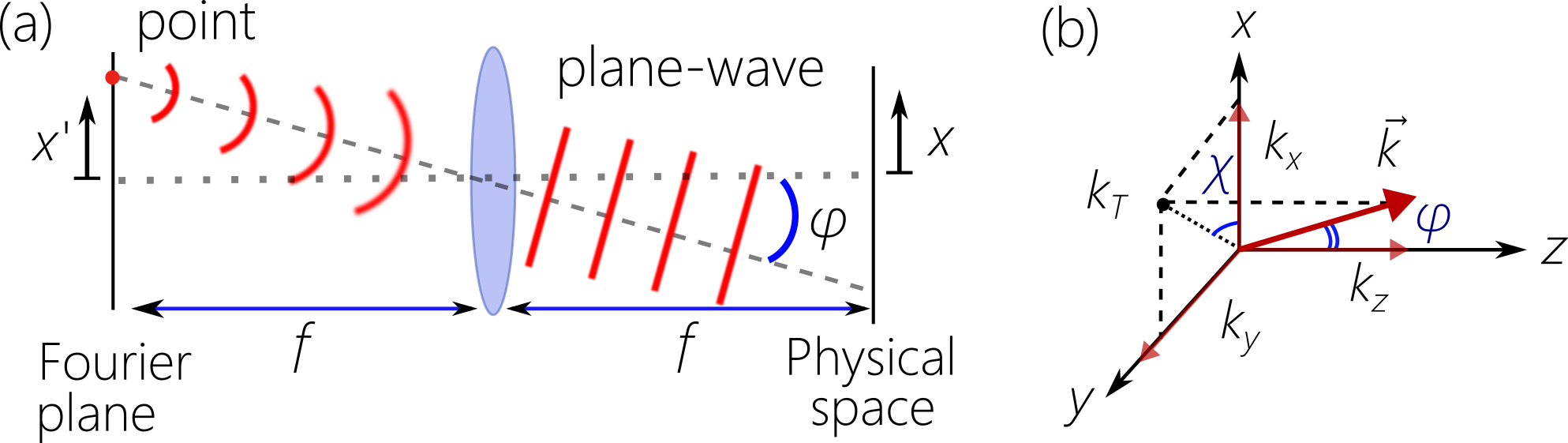}
    \caption{(a) A $2f$ system converts a point in the back focal plane to a plane wave. (b) Geometry of the wave vector in the Fourier plane.}
    \label{fig:BasicFourierOptics}
\end{figure}

Traditional treatments of Fourier optics are typically concerned with monochromatic light, or at least the temporal bandwidth does not play an essential role. Consider an optical field at frequency $\omega_{\mathrm{o}}$ in which the components of the wave vector satisfy the condition $k_{x}^{2}+k_{y}^{2}+k_{z}^{2}\!=\!k_{\mathrm{o}}^{2}$, where $k_{x}$ and $k_{y}$ are the transverse components along the $x$ and $y$ axes, respectively, $k_{z}$ is the axial component along $z$, $k_{\mathrm{o}}\!=\!\tfrac{\omega_{\mathrm{o}}}{c}$ is the wave number associated with $\omega_{\mathrm{o}}$, and $c$ is the speed of light in vacuum. It is customary in Fourier optics to refer to the transverse wave-vector components $(k_{x},k_{y})$ as the spatial frequencies. Because we are interested here in extending the concepts of traditional (spatial) Fourier optics to the more generalized spatiotemporal Fourier optics, we will refer to $\omega_{\mathrm{o}}$ as the \textit{temporal frequency} to symmetrize the treatment of the spatial and temporal degrees of freedom. The central construct of Fourier optics is the $2f$ optical system depicted in Fig.~\ref{fig:BasicFourierOptics}(a), whereupon a point source in the back focal plane of a lens is mapped to a plane wave after the lens (and vice versa) \cite{SalehBook07}. Indeed, if the monochromatic field in the back focal plane takes the ideal form $\psi'(x',y')\!\propto\!\delta(x'-x_{\mathrm{o}},y'-y_{\mathrm{o}})$, then the field in the focal plane is $\psi(x,y)\!\propto\!e^{i\frac{k_{\mathrm{o}}}{f}(x_{\mathrm{o}}x+y_{\mathrm{o}}y)}\!=\!e^{i(k_{x}x+k_{y}y)}$, where $f$ is the lens focal length. We therefore have the following association between position in the Fourier plane and spatial frequency:
\begin{equation}\label{eq:SpatialFrequencies}
k_{x}=\frac{k_{\mathrm{o}}}{f}x',\;\;\;k_{y}=\frac{k_{\mathrm{o}}}{f}y'.
\end{equation}

The central idea of Fourier optics is that this association between positions in Fourier space and spatial frequencies in the field produced in physical space leads to the following equivalency: preparing the complex field $\psi'(x',y')$ in the Fourier plane $(x',y')$ is equivalent to preparing the complex spatial spectrum of the field produced in physical space in the front focal plane $(x,y)$ of the $2f$ system in Fig.~\ref{fig:BasicFourierOptics}(a). Consider a monochromatic field in physical space $E(x,y,z;t)\!=\!e^{i(k_{\mathrm{o}}z-\omega_{\mathrm{o}}t)}\psi(x,y,z)$, where $z$ is measured from the focal plane, and the envelope is given by:
\begin{equation}
\psi(x,y,z)=\iint\!dk_{x}dk_{y}\widetilde{\psi}(k_{x},k_{y})e^{i(k_{x}x+k_{y}y)}e^{i(k_{z}-k_{\mathrm{o}})z},
\end{equation}
where $k_{z}\!=\!\sqrt{k_{\mathrm{o}}^{2}-k_{x}^{2}-k_{y}^{2}}$, and the spatial spectrum $\widetilde{\psi}(k_{x},k_{y})$ is the Fourier transform of $\psi(x,y,0)$. The goal in Fourier optics is to prepare a field distribution in the Fourier plane of the form:
\begin{equation}
\psi'(x',y')\propto\widetilde{\psi}\left(\frac{k_{\mathrm{o}}}{f}x',\frac{k_{\mathrm{o}}}{f}y'\right).
\end{equation}

\subsection{Fourier optics in terms of propagation angles}

When dealing with the pulsed scenario below, where $k_{\mathrm{o}}$ is no longer fixed, it will be more useful to deal with propagation \textit{angles} rather than transverse wave numbers or spatial frequencies. The reason can be understood by reference to Eq.~\ref{eq:SpatialFrequencies} where the position $(x',y')$ is proportional to the spatial frequency $(k_{x},k_{y})$, but the proportionality involves the temporal frequency ($k_{\mathrm{o}}\!=\!\tfrac{\omega_{\mathrm{o}}}{c}$). In the case of a monochromatic field, this does not pose a difficulty and represents a fixed scaling factor. However, when the field has a finite temporal bandwidth $\Delta\omega$, the association between position in the Fourier plane and the spatial frequency is no longer unique. Indeed, two temporal frequencies $\omega_{1}$ and $\omega_{2}$ at the \textit{same} point $(x',y')$ in the Fourier plane give rise to two \textit{different} spatial frequencies: $(k_{x1},k_{y1})\!=\!\tfrac{\omega_{1}}{cf}(x',y')$ produced for $\omega_{1}$, and $(k_{x2},k_{y2})\!=\!\tfrac{\omega_{2}}{cf}(x',y')$ for $\omega_{2}$. Consequently, a fixed spatial frequency $(k_{x},k_{y})$ can occupy the entire Fourier plane as long as the temporal frequency $\omega$ changes to maintain $(k_{x},k_{y})=(\tfrac{\omega}{cf}x',\tfrac{\omega}{cf}y')$, in which case a single spatial frequency is excited despite the extended Fourier domain illuminated. 

To avoid this difficulty, we associate \textit{positions} in the Fourier plane with \textit{propagation angles} rather than \textit{spatial frequencies}. The geometry is depicted in Fig.~\ref{fig:BasicFourierOptics}(b). The wave vector $\vec{k}$ with magnitude $k\!=\!\tfrac{\omega}{c}$ makes an angle $\varphi$ with the $z$-axis, so that $k_{z}\!=\!k\cos{\varphi}$, whereas the transverse wave vector components are given by $k_{x}=k_{\mathrm{T}}\cos\chi=k\cos\chi\sin\varphi$ and $k_{y}=k_{\mathrm{T}}\sin\chi=k\sin\chi\sin\varphi$, where $k_{\mathrm{T}}\!=\!k\sin\varphi\!=\!\sqrt{k_{\mathrm{o}}^{2}-k_{z}^{2}}$ and $\tan\chi=k_{y}/k_{x}$; see Fig.~\ref{fig:BasicFourierOptics}(b). Substituting in Eq.~\ref{eq:SpatialFrequencies}, a position $(x',y')$ in the Fourier plane is associated with the propagation angles $\varphi$ and $\chi$ according to:
\begin{equation}\label{eq:Angles}
x'=f\cos\chi\sin\varphi,\;\;\;y'=f\sin\chi\sin\varphi,
\end{equation}
where $\sin\varphi\!=\!\tfrac{1}{f}\sqrt{x'^{2}+y'^{2}}$ and $\tan\chi\!=\!y'/x'$.

These relationships are independent of $\omega$, and the propagation angles are uniquely determined by the position in the Fourier plane. In other words, \textit{any} frequency $\omega$ located at the position $(x',y')$ in the Fourier plane will produce a plane wave with the \textit{same} propagation angles $\varphi$ and $\chi$ (albeit associated with different spatial frequencies). Relying on propagation angles, therefore offers convenient variables in the Fourier plane, which is critical for spatiotemporal Fourier optics.

\subsection{`Diffraction-free' monochromatic fields}

In light of our interest here in propagation-invariant pulsed beams, we consider two elementary monochromatic field configurations that are 'diffraction-free': the cosine wave, which makes use of one transverse dimension (1D), and the Bessel beam that makes use of two transverse dimensions (2D).

\begin{figure}[t!]
    \centering
    \includegraphics[width=8.6cm]{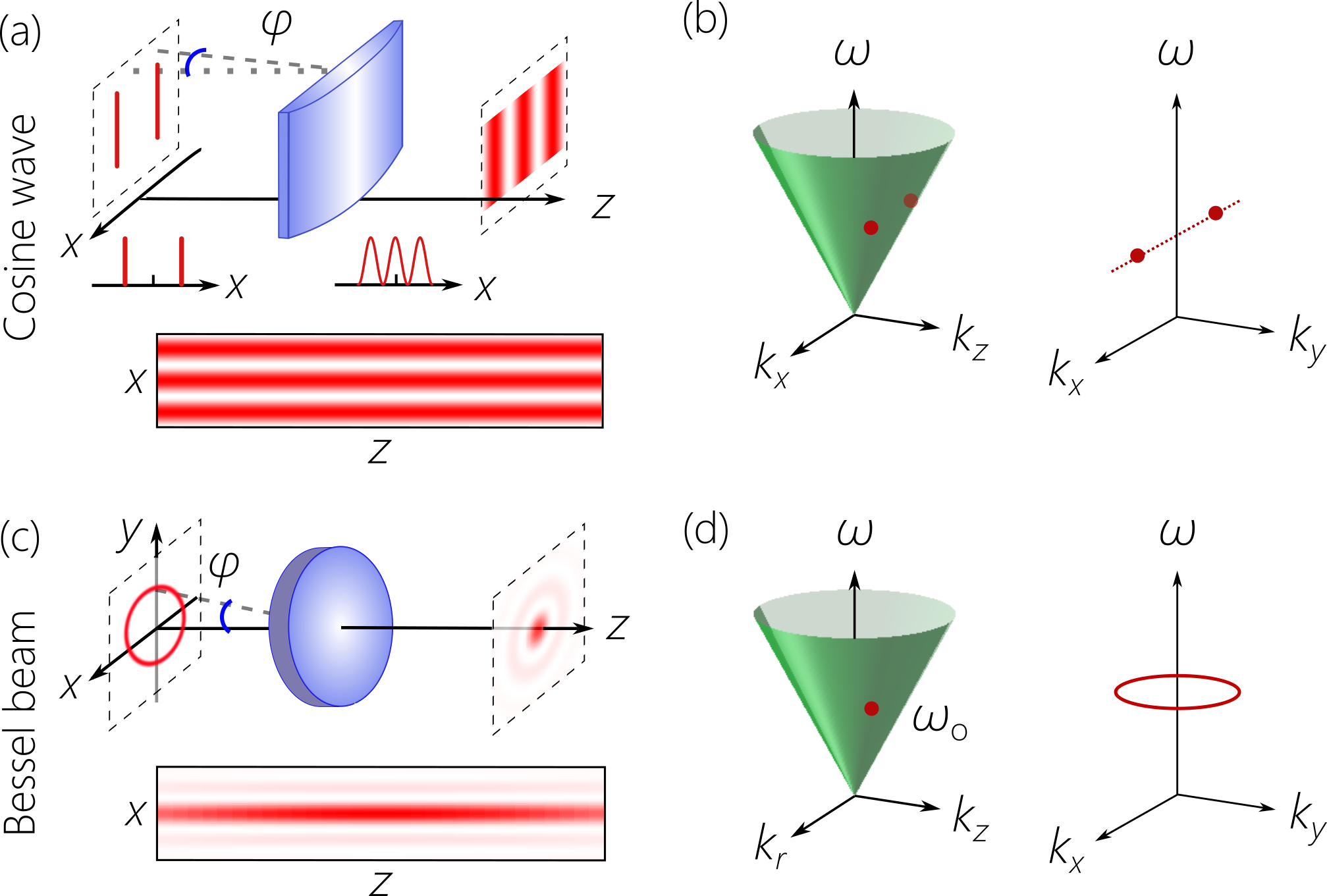}
    \caption{`Diffraction-free' monochromatic fields having 1D and 2D spatial Fourier spectra. (a) A cylindrical lens converts a fixed spatial frequency $k_{x}\!=\!|k_{\mathrm{T}}|$ into a diffraction-free cosine wave. (b) Representation of the cosine wave on the surface of the reduced light-cone $k_{x}^{2}+k_{z}^{2}\!=\!(\tfrac{\omega}{c})^{2}$ in the left panel, and in the reduced $(k_{x},k_{y},\tfrac{\omega}{c})$-space in the right panel. (c) A spherical lens converts a fixed spatial frequency $k_{r}\!=\!k_{\mathrm{T}}$ into a Bessel beam. (d) Representation of the Bessel beam on the surface of the light-cone $k_{r}^{2}+k_{z}^{2}\!=\!(\tfrac{\omega}{c})^{2}$ in the left panel, and in the reduced $(k_{x},k_{y},\tfrac{\omega}{c})$-space in the right panel.}
    \label{fig:DiffractionFreeBeams}
\end{figure}

\subsubsection{Cosine waves in one transverse dimension}

We first consider a scenario restricted to one transverse spatial dimension $x$ (the field is uniform along $y$), so that a cylindrical lens suffices [Fig.~\ref{fig:DiffractionFreeBeams}(a)]. The field is now $E(x,y,z;t)\!=\!e^{i(k_{\mathrm{o}}z-\omega_{\mathrm{o}}t)}\psi(x,z)$, and the envelope is given by: $\psi(x,z)\!=\!\int\!dk_{x}\widetilde{\psi}(k_{x})e^{ik_{x}x}e^{i(k_{z}-k_{\mathrm{o}})z}$. Illuminating a line in the Fourier plane at position $x'\!=\!x_{\mathrm{o}}$ with the temporal frequency $\omega_{\mathrm{o}}$ yields a plane wave with spatial frequency $k_{x}\!=\!k_{\mathrm{o}}\tfrac{x_{\mathrm{o}}}{f}\!=\!k_{\mathrm{c}}$. A symmetrically illuminated line at $-x_{\mathrm{o}}$ maps to a plane wave with $k_{x}\!=\!-k_{\mathrm{c}}$. Therefore, the pair of lines, which represent the locus of constant spatial frequency $|k_{\mathrm{T}}|$, yields a cosine wave: $E(x,z;t)\propto\cos(k_{\mathrm{c}}x)e^{i\beta z}e^{-i\omega_{\mathrm{o}}t}$, where $\beta\!=\!\sqrt{k_{\mathrm{o}}^{2}-k_{\mathrm{c}}^{2}}\!=\!k_{\mathrm{o}}\sqrt{1-(\tfrac{x_{\mathrm{o}}}{f})^{2}}$ is a fixed axial wave number. The same cosine wave is obtained after replacing the cylindrical lens with a spherical lens, and replacing the lines $x'\!=\!\pm x_{\mathrm{o}}$ extended along $y'$ with two points at $(x',y')\!=\!(\pm x_{\mathrm{o}},0)$.  

The field produced is separable with respect to the transverse and axial coordinates $x$ and $z$, respectively, such that $|E(x,z;t)|\!=\!|E(x,0;t)|$. Such a monochromatic optical field is typically referred to as `diffraction-free' \cite{Durnin87PRL,Turunen10PO}; see Fig.~\ref{fig:DiffractionFreeBeams}(a). It is straightforward to demonstrate that the cosine wave is the only diffraction-free monochromatic 1D beam \cite{Yessenov22AOP}. The only exception is the 1D Airy beam whose transverse profile is fixed with propagation, but whose peak travels along a parabolic trajectory rather than a straight line \cite{Siviloglou07OL,Hu13PRA,Bongiovanni16OE,Efremidis19Optica}.

It is useful to consider the representation of the cosine wave on the surface of the light-cone $k_{x}^{2}+k_{y}^{2}+k_{z}^{2}\!=\!(\tfrac{\omega}{c})^{2}$. This is a conceptual and visualization tool that we rely on heavily here and has been used extensively in the literature \cite{Donnelly93PRSLA,Yessenov22AOP}. Because this 4D hyper-cone cannot be represented graphically in 3D, we make use instead of the reduced light-cone $k_{x}^{2}+k_{z}^{2}\!=\!(\tfrac{\omega}{c})^{2}$ since $k_{y}\!=\!0$ in the 1D Fourier spatial spectrum considered here. The cosine wave is represented by two points along the circle located at the intersection of the light-cone with the horizontal iso-frequency plane $\omega\!=\!\omega_{\mathrm{o}}$. The two points are placed symmetrically with respect to the $k_{z}$-axis; see Fig.~\ref{fig:DiffractionFreeBeams}(b).

\subsubsection{Bessel beam in two transverse dimensions}

We consider next the equivalent scenario in 2D Fourier space to the cosine wave in 1D Fourier space, in which case the locus of a constant spatial frequency in the Fourier plane is a circle of radius $k_{\mathrm{B}}$. Converting the angular spectrum into polar coordinates, the envelope $\psi(r,\varphi_{\mathrm{T}},z)$ is:
\begin{equation}
\psi(r,\varphi_{\mathrm{T}},z)=\!\int_{0}^{k_{\mathrm{o}}}\!\!k_{r}dk_{r}\int_{-\pi}^{\pi}\!\!d\chi\,\widetilde{\psi}(k_{r},\chi)e^{ik_{r}r\cos(\varphi_{\mathrm{T}}-\chi)}e^{i(k_{z}-k_{\mathrm{o}})z},
\end{equation}
where $k_{r}^{2}\!=\!k_{x}^{2}+k_{y}^{2}$, $\tan\chi\!=\!\tfrac{k_{y}}{k_{x}}$, $r^{2}\!=\!x^{2}+y^{2}$, and $\tan\varphi_{\mathrm{T}}\!=\!\tfrac{y}{x}$. The spatial spectrum $\widetilde{\psi}(k_{r},\chi)$ can be decomposed along $\chi$ in terms of orbital angular momentum modes $\ell$, $\widetilde{\psi}(k_{r},\chi)\!=\!\tfrac{1}{2\pi}\sum_{\ell\!=\!-\infty}^{\infty}\widetilde{\psi}_{\ell}(k_{r})e^{i\ell\chi}$, so that the envelope can be written as:
\begin{equation}
\psi(r,\varphi_{\mathrm{T}},z)=\sum_{\ell=-\infty}^{\infty}e^{i\ell\varphi_{\mathrm{T}}}\int_{0}^{k_{\mathrm{o}}}\!dk_{r}\,k_{r}\widetilde{\psi}_{\ell}(k_{r})J_{\ell}(k_{r}r)e^{i(k_{z}-k_{\mathrm{o}})z},
\end{equation}
where we have made use of the identity $2\pi J_{\ell}(x)\!=\!\int_{-\pi}^{\pi}\!dy\,e^{i(x\sin{y}-\ell y)}$. For an azimuthally symmetric field that is independent of $\varphi_{\mathrm{T}}$, we set $\ell\!=\!0$, and obtain:
\begin{equation}\label{eq:MonochromaticAzimuthSymmetric}
\psi(r,z)=\int_{0}^{k_{\mathrm{o}}}\!dk_{r}\widetilde{\psi}(k_{r})k_{r}J_{0}(k_{r}r)e^{i(k_{z}-k_{\mathrm{o}})z}.
\end{equation}
In other words, any azimuthally symmetric monochromatic beam is the appropriately weighted superposition of Bessel beams of different transverse wave numbers $k_{r}$.

The simplest case for Eq.~\ref{eq:MonochromaticAzimuthSymmetric} is to restrict $\widetilde{\psi}(k_{r})$ to a single constant transverse wave number (or radial spatial frequency), $\widetilde{\psi}(k_{r})\!=\!\tfrac{1}{k_{\mathrm{B}}}\delta(k_{r}-k_{\mathrm{B}})$, in which case:
\begin{equation}
E(r,z;t)\propto J_{0}(k_{\mathrm{B}}r)e^{i\beta z}e^{-i\omega_{\mathrm{o}}t},
\end{equation}
where $\beta\!=\!\sqrt{k_{\mathrm{o}}^{2}-k_{\mathrm{B}}^{2}}$ is a constant axial wave number. The spatial spectrum in the Fourier plane is located on a circle [Fig.~\ref{fig:DiffractionFreeBeams}(c)]. Each point on this circle maps to a plane wave, and the multiplicity of points along the circle maps to plane waves traveling at the same angle with the $z$-axis, thus forming a cone. Similarly to the cosine wave, this field is diffraction-free $|E(r,z;t)|\!=\!|E(r,0;t)|$, and is separable with respect to the transverse and axial coordinates [Fig.~\ref{fig:DiffractionFreeBeams}(c)]. 

The monochromatic Bessel beam is represented in Fig.~\ref{fig:DiffractionFreeBeams}(d) in one of two spectral spaces: (1) in $(k_{r},k_{z},\tfrac{\omega}{c})$-space it is represented on the surface of the reduced light-cone $k_{r}^{2}+k_{z}^{2}\!=\!(\tfrac{\omega}{c})^{2}$ by a point $(k_{\mathrm{B}},\beta,\tfrac{\omega_{\mathrm{o}}}{c})$; and (2) in the reduced space $(k_{x},k_{y},\tfrac{\omega}{c})$ where the spatial Fourier spectrum associated with each temporal frequency lies in a plane at a different height, the spatial spectrum of a monochromatic Bessel beam is a circle of radius $k_{\mathrm{B}}$ in the plane $\omega\!=\!\omega_{\mathrm{o}}$.

Both the cosine and Bessel beams have infinite power at any $z$: $\int\!dx|E(x,z)|^{2}\rightarrow\infty$ for the cosine wave, and $\int\!rdr|E(r,z)|^{2}\!\rightarrow\!\infty$ for the Bessel beam. In practice, however, the ideal spatial spectrum cannot be realized. For example, only a finite-thickness annulus in $(k_{x},k_{y},\tfrac{\omega}{c})$-space can be achieved for a Bessel beam rather than a geometric circle. This spatial uncertainty sets a finite limit on the propagation distance of realistic Bessel beams \cite{McGloin05CP,Mazilu10LPR}.

\section{Spatiotemporal Fourier synthesis}

We now consider the case of spatiotemporal Fourier optics; that is, spectral shaping in the joint spatial and temporal degrees of freedom of a field of finite temporal bandwidth $\Delta\omega$. This applies in principle to fields that are spectrally coherent (pulsed fields) or incoherent (continuous-wave fields). We focus here on the former case of pulsed optical fields. In this scenario, the advantage of relying on Eq.~\ref{eq:Angles} (where positions in the Fourier plane are mapped to \textit{propagation angles}) rather than Eq.~\ref{eq:SpatialFrequencies} (where positions in the Fourier plane are mapped to \textit{spatial frequencies}) will become clear.

The basic equations used above for the monochromatic scenario can be readily extended to the case of pulsed fields $E(x,y,z;t)\!=\!e^{i(k_{\mathrm{o}}z-\omega_{\mathrm{o}}t)}\psi(x,y,z;t)$, where:
\begin{equation}
\psi(x,y,z;t)\!=\!\iiint\!dk_{x}dk_{y}d\Omega\;\widetilde{\psi}(k_{x},k_{y},\Omega)e^{i(k_{x}x+k_{y}y+(k_{z}-k_{\mathrm{o}})z-\Omega t)};
\end{equation}
here $\Omega\!=\!\omega-\omega_{\mathrm{o}}$, and $\omega_{\mathrm{o}}$ is a fixed temporal frequency. We will focus on the reduced problem of azimuthally symmetric fields $\widetilde{\psi}(k_{x},k_{y},\Omega)\!\rightarrow\!\widetilde{\psi}(k_{r},\Omega)$, which reduces the dimensionality of the spatiotemporal spectrum. Following the same formulation used above for a monochromatic field, a pulsed azimuthally symmetric field takes, in general, the form:
\begin{equation}
\psi(r,z;t)=\iint\!k_{r}dk_{r}d\Omega\;\widetilde{\psi}(k_{r},\Omega)J_{0}(k_{r}r)e^{i(k_{z}-k_{\mathrm{o}})z}e^{-i\Omega t}.
\end{equation}

\begin{figure}[t!]
    \centering
    \includegraphics[width=8.6cm]{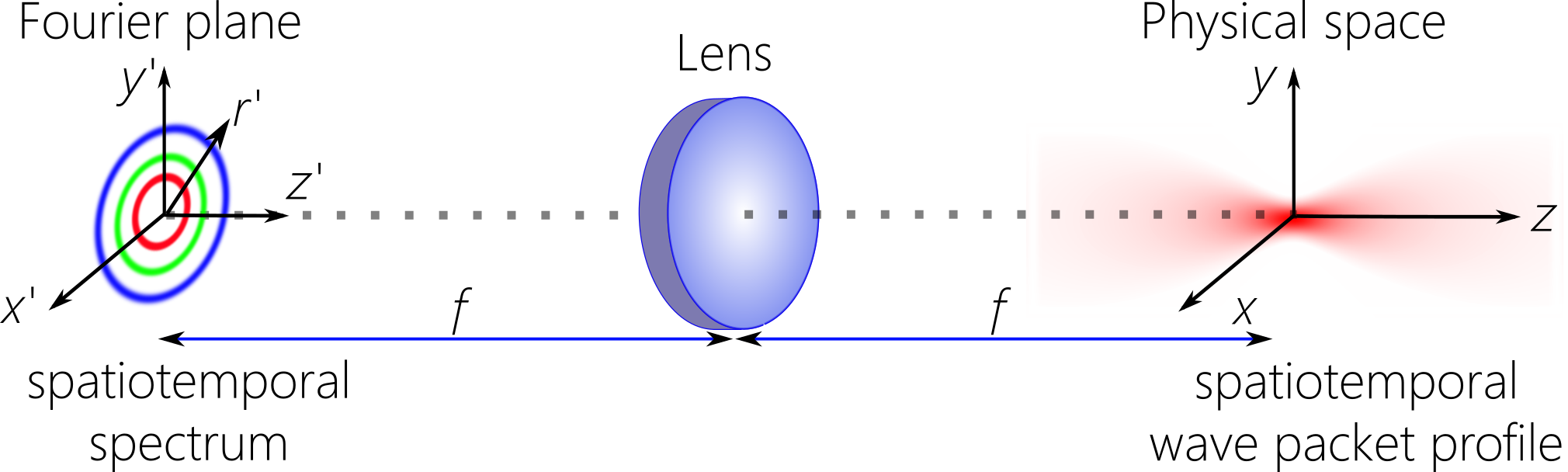}
    \caption{Schematic of the spatiotemporal spectrum in the Fourier plane of the sub-class of spatiotemporally structured fields dealt with here. Each temporal frequency $\omega$ occupies an annulus of radius $r'(\omega)$, which maps to a radial spatial frequency $k_{r}$. Preparing this field structure constitutes the spatiotemporal spectral synthesis step.}
    \label{fig:SpatiotemporalSpectrum}
\end{figure}

Even after this simplification, exploring the full space of spatiotemporal spectra $\widetilde{\psi}(k_{r},\Omega)$ remains a daunting prospect. We therefore focus on a subset of spatiotemporally structured fields in which each temporal frequency $\Omega$ is associated with a single spatial frequency $k_{r}$. In this case, the field in the Fourier plane takes the form of a sequence of rings or annuluses, each of which has a different temporal frequency [Fig.~\ref{fig:SpatiotemporalSpectrum}]. The radius of the ring associated with $\omega$ is $r'(\omega)$, the propagation angle after the $2f$ lens is $\varphi(\omega)$ as determined from $r'(\omega)=f\sin\left\{\varphi(\omega)\right\}$, and the associated spatial frequency is:
\begin{equation}
k_{r}(\omega)=\frac{k}{f}r'(\omega)=k\sin\left\{\varphi(\omega)\right\}.
\end{equation}

The central problem in spatiotemporal Fourier optics is the determination of the radial distribution $r'(\omega)$ in the Fourier plane that yields the desired angular spectral profile $\varphi(\omega)$, and thus the frequency-dependent spatial frequency $k_{r}(\omega)$, which is needed to construct the target spatiotemporally structured field in space and time. The successful practical implementation of spatiotemporal Fourier optics therefore requires, in general, the ability to locate each frequency $\omega$ at a prescribed radius $r'(\omega)$ in the Fourier domain [Fig.~\ref{fig:SpatiotemporalSpectrum}].

When dealing with optical fields in which the temporal spectrum is spatially resolved along one transverse dimension, as in the case with prisms and diffraction gratings, the frequency-dependence of the propagation angle is usually referred to as angular dispersion (AD) \cite{Hebling96OQE,Fulop10Review,Torres10AOP,Hall21OL,Hall22OEConsequences}. For circularly symmetric fields [Fig.~\ref{fig:SpatiotemporalSpectrum}], $\varphi(\omega)$ is usually referred to as the `cone angle', and the frequency dependence of $\varphi$ can be termed conical-AD. In the following, we explore several case studies of interest in which pulsed beams whose unique characteristics stem from their particular spatiotemporal spectral distribution in the Fourier plane.

\section{Case I: Pulsed Bessel beams}

Consider the wave packet known as the pulsed Bessel beam \cite{Hu02JOSAA,Lu03JOSAA,Hall25PRA}, which is defined as the pulsed beam for which $k_{r}(\omega)\!=\!k_{\mathrm{B}}$; i.e., the spatiotemporal spectrum contains a single transverse wave number or spatial frequency $k_{\mathrm{B}}$, which is associated with all the temporal frequencies \cite{Turunen10PO}.

\begin{figure}[t!]
    \centering
    \includegraphics[width=8.6cm]{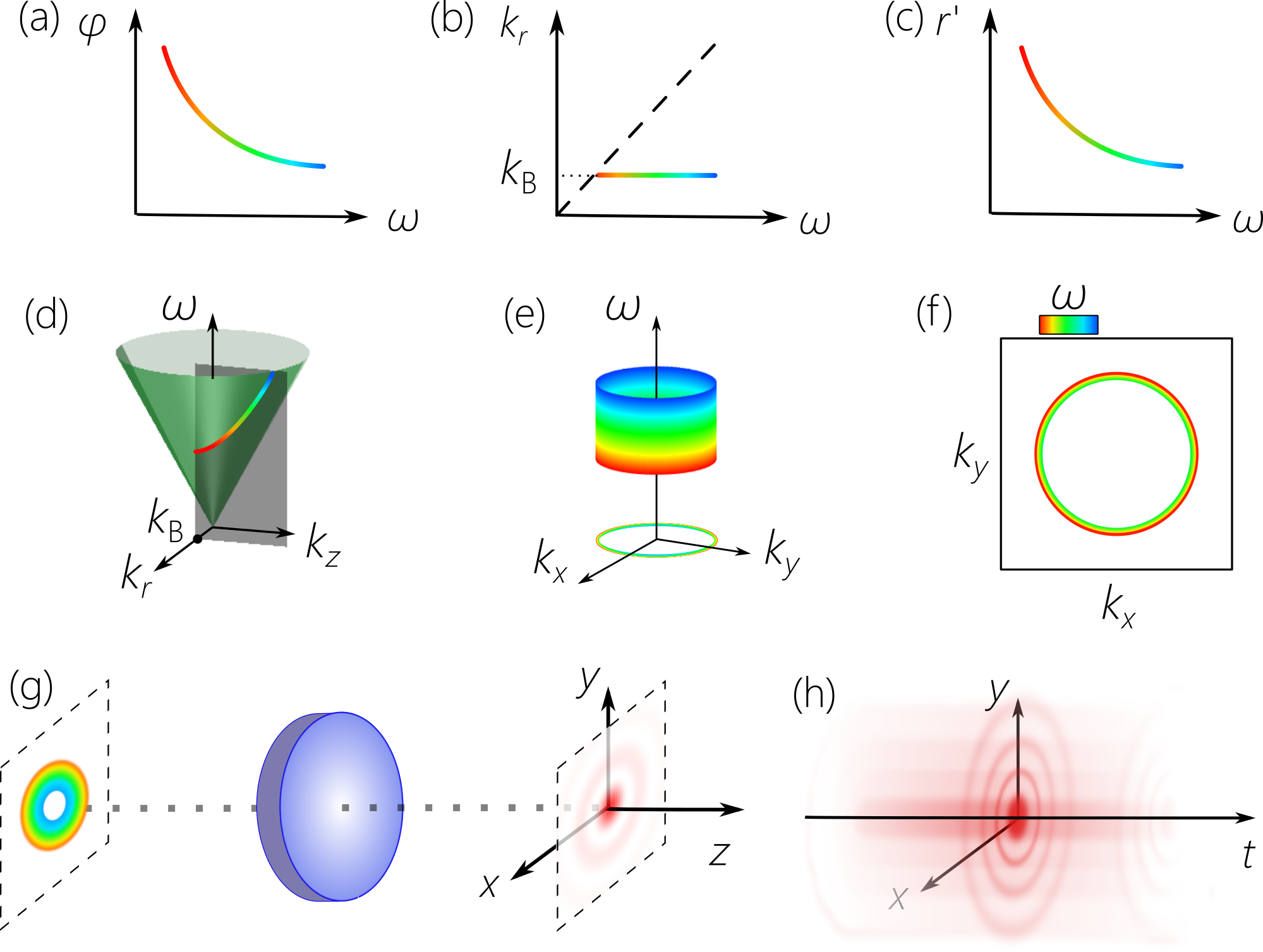}
    \caption{Spatiotemporal structure of a pulsed Bessel beam. (a) The spectral AD profile $\varphi(\omega)\!\propto\!\frac{1}{\omega}$ (in the small angle limit). (b) The spatiotemporal spectrum is associated with a constant radial spatial frequency $k_{r}\!=\!k_{\mathrm{B}}$. The dashed line is the light-line $k_{z}\!=\!\tfrac{\omega}{c}$ ($k_{z}=0$; evanescent waves lie above the light-line). (c) The spectral distribution in the Fourier plane, with $r(\omega)\!\propto\!\tfrac{1}{\omega}$. (d) The spatiotemporal spectrum of a pulsed Bessel beam on the surface of the light-cone along the hyperbola at its intersection with an iso-$k_{r}$ plane. (e) The spatiotemporal spectrum in $(k_{x},k_{y},\tfrac{\omega}{c})$-space is a cylinder of radius $k_{\mathrm{B}}$. (f) Projection of the spectrum in (e) onto the $(k_{x},k_{y})$-plane. (g) The field distribution in the back and front focal planes of a $2f$ system for a pulsed Bessel beam. In the Fourier plane, the field takes the form of a disk with radially arranged wavelengths where $r(\omega)\!\propto\!\tfrac{1}{\omega}$. (h) In physical space, the field is separable with respect to the spatial and temporal degrees-of-freedom.}
    \label{fig:PulsedBessel}
\end{figure}

One may expect the spectrum to comprise a single annulus in the Fourier plane because of the constant transverse wave number $k_{r}=k_{\mathrm{B}}$ in analogy with the Bessel beam. However, because positions in the Fourier plane do \textit{not} map uniquely to spatial frequencies when the field has finite temporal bandwidth $\Delta\omega$, the spatiotemporal spectrum instead takes the form shown in Fig.~\ref{fig:PulsedBessel}. The propagation angle $\varphi(\omega)$ is determined from $\sin\{\varphi(\omega)\}\!=\!\tfrac{k_{\mathrm{B}}}{k}\!\propto\!\tfrac{1}{\omega}$ [Fig.~\ref{fig:PulsedBessel}(a)], from which we obtain the spatial distribution of the temporal frequencies in the Fourier plane $r'(\omega)\!=\!\tfrac{fk_{\mathrm{B}}}{k}\!\propto\!\tfrac{1}{\omega}$ [Fig.~\ref{fig:PulsedBessel}(c)]. Shorter wavelengths (higher temporal frequencies) are located at smaller radii. Although the spatial frequency is constant $k_{r}(\omega)\!=\!k_{\mathrm{B}}$ [Fig.~\ref{fig:PulsedBessel}(b)], the spatiotemporal spectrum nevertheless extends across the Fourier plane.

It is instructive to examine the spectral support of the pulsed Bessel beam on the light-cone surface. In $(k_{r},k_{z},\tfrac{\omega}{c})$-space, the spectrum is the hyperbola at the intersection of the light-cone with a vertical iso-$k_{r}$ plane, $k_{z}^{2}\!=\!(\tfrac{\omega}{c})^{2}-k_{\mathrm{B}}^{2}$ [Fig.~\ref{fig:PulsedBessel}(d)]. In $(k_{x},k_{y},\tfrac{\omega}{c})$-space, the spectrum takes is a cylinder with radius $k_{\mathrm{B}}$ [Fig.~\ref{fig:PulsedBessel}(e,f)]. This spectral surface is spatiotemporally separable, and the spatiotemporal field structure is consequently separable into a product of spatial and temporal envelopes [Fig.~\ref{fig:PulsedBessel}(g,h)]: $\psi(r,z;t)=J_{0}(k_{\mathrm{B}}r)\psi_{t}(z;t)$, where the temporal envelope $\psi_{t}(z;t)$ travels at a group velocity $\widetilde{v}\!=\!c$, but undergoes anomalous group-velocity dispersion (GVD) in free space:
\begin{equation}
\psi_{t}(z;t)\approx\int\!d\Omega\,\widetilde{\psi}(\Omega)e^{-i\Omega(t-z/c)}e^{-i\frac{k_{\mathrm{B}}^{2}}{2k_{\mathrm{o}}}\tfrac{\Omega^{2}}{\omega_{\mathrm{o}}^{2}}z}.
\end{equation}
The fact that a pulsed Bessel beam is dispersive can be derived immediately from the frequency-dependence of the axial wave number $k_{z}(\omega)=\sqrt{(\tfrac{\omega}{c})^{2}-k_{\mathrm{B}}^{2}}$, from which $\tfrac{d^{2}k_{z}}{d\omega^{2}}\big|_{\omega_{\mathrm{o}}}=-\tfrac{1}{c^{2}}\tfrac{k_{\mathrm{B}}^{2}}{(k_{\mathrm{o}}^{2}-k_{\mathrm{B}}^{2})^{3/2}}\approx-\tfrac{k_{\mathrm{B}}^{2}}{c^{2}k_{\mathrm{o}}^{3}}<0$, corresponding to anomalous GVD. A pulsed Bessel beam is \textit{not} propagation invariant in free space. Although the spatial profile takes the form of a diffraction-free Bessel beam, the pulse profile undergoes dispersive broadening [Fig.~\ref{fig:PulsedBessel}(g,h)]. Such pulsed Bessel beams can be used for GVD cancellation in media characterized by normal GVD, whereupon they become propagation invariant \cite{Turunen10PO,Hall25PRA}.

Sending a pulse into conventional Bessel beam generators typically produces X-waves (see next Section) rather than pulsed Bessel beams. A pulsed Bessel beam can be produced by imposing the Bessel beam profile on a plane-wave pulse in physical space, or by synthesizing the spatiotemporal spectrum shown in Fig.~\ref{fig:PulsedBessel}. Nevertheless, such a beam requires a large numerical aperture \cite{Malaguti09PRA} to propagate invariantly in a dispersive medium \cite{Hall25PRA}. Note that the properties of the pulsed Bessel beam are independent of the particular form of the pulse envelope $\psi_{t}(t)$. Utilizing an Airy pulse instead, the so-called Bessel-Airy wave packet \cite{Chong10NP}, alleviate this numerical aperture requirement.

\begin{figure}[t!]
    \centering
    \includegraphics[width=8.6cm]{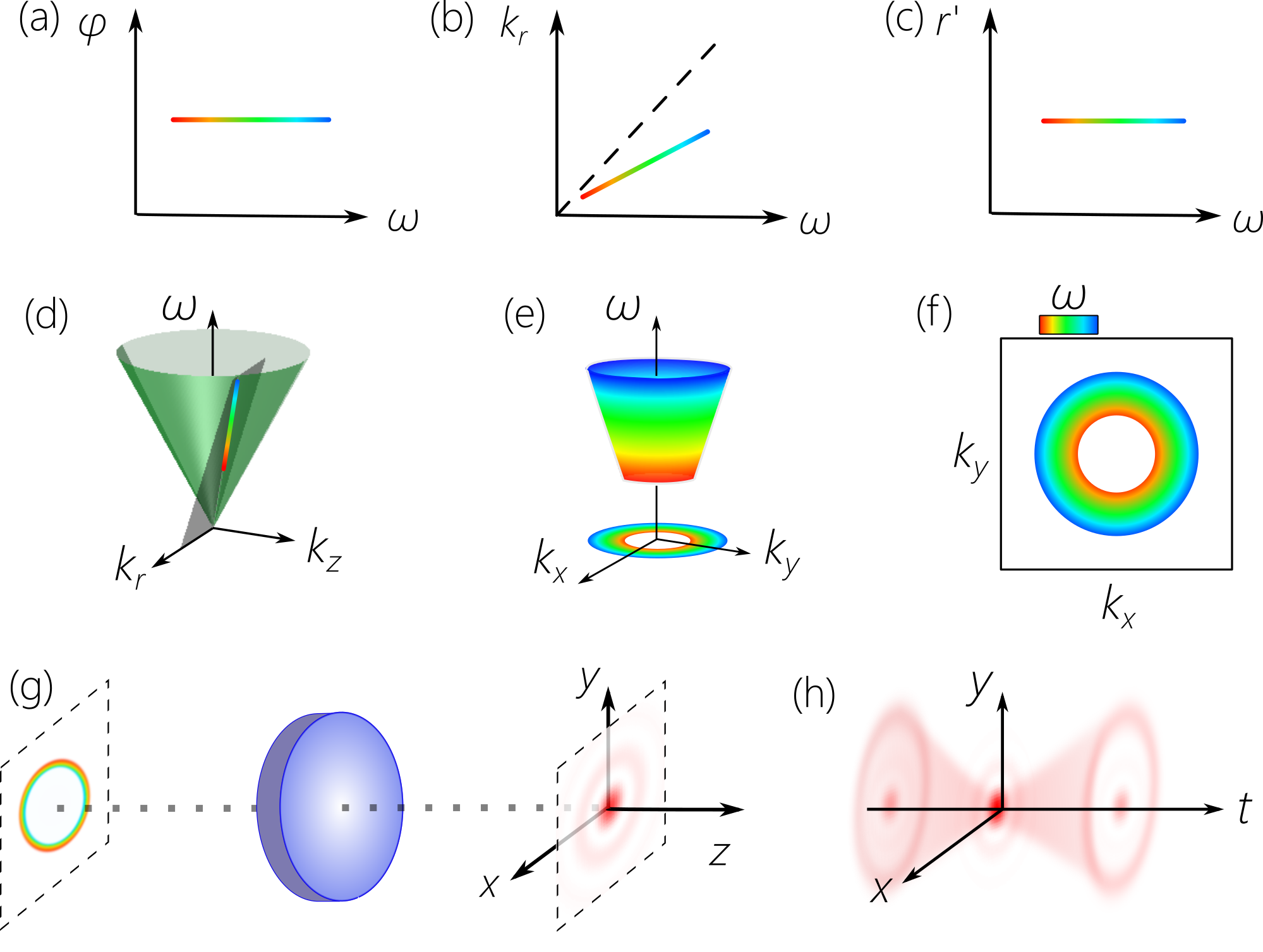}
    \caption{Spatiotemporal structure of an X-wave (following Fig.~\ref{fig:PulsedBessel}). (a) The spectral AD profile $\varphi(\omega)\!=\!\varphi_{\mathrm{o}}$. (b) The spatiotemporal spectrum is a straight line, (c) but all the frequencies are located at the same radius. (d) The spectral support is the line at the intersection of the light-cone with a tilted plane that includes the $k_{r}$-axis and makes an angle $\theta$ with the $k_{z}$-axis. (e) The spatiotemporal spectrum in $(k_{x},k_{y},\tfrac{\omega}{c})$-space is the surface of a cone. (f) Projection of the spectrum in (e) onto the $(k_{x},k_{y})$-plane. (g) In the Fourier plane, the field is an annulus. (h) In physical space, the cylindrically symmetric X-wave is X-shaped in a meridional plane.}
    \label{fig:XwaveCones}
\end{figure}

\section{Case II: X-waves}

The X-wave is a cylindrically symmetric pulsed beam that propagates in free space without diffraction or dispersion at a superluminal group velocity $\widetilde{v}\!>\!c$. The crucial feature of an X-wave is that the propagation angle is fixed $\varphi(\omega)\!=\!\varphi_{\mathrm{o}}$ [Fig.~\ref{fig:XwaveCones}(a)], so that $k_{r}\!=\!\tfrac{\omega}{c}\sin{\varphi_{\mathrm{o}}}\!\propto\!\omega$ [Fig.~\ref{fig:XwaveCones}(b)]. Although the \textit{radial spatial frequency} changes with temporal frequency, the \textit{propagation angle} is fixed, so the temporal frequencies are co-located at the radius $r'(\omega)\!=\!f\sin{\varphi_{\mathrm{o}}}$ in the Fourier domain [Fig.~\ref{fig:XwaveCones}(c)]. 

Because the radial spatial frequency $k_{r}$ is linear in $\omega$, the spectral support of an X-wave on the surface of the light-cone in $(k_{r},k_{z},\tfrac{\omega}{c})$-space is at its intersection with a spectral plane that passes through the origin, is parallel to the $k_{r}$-axis, and makes an angle $\theta$ with the $k_{z}$-axis (which we refer to as the spectral tilt angle). The field associated with an X-wave is thus [Fig.~\ref{fig:XwaveCones}(g)]:
\begin{equation}\label{eq:XwaveField}
E(r,z;t)=\int\!d\omega\,\widetilde{\psi}(\omega)e^{-i\omega(t-z/\widetilde{v})}=E(r,0;t-z/\widetilde{v}).
\end{equation}
This is a propagation-invariant wave packet traveling at a group velocity $\widetilde{v}\!=\!c\tan{\theta}$, where $\tan{\theta}\!=\!\sec{\varphi_{\mathrm{o}}}$. The carrier term is absent because the phase and group velocities are equal. 

X-waves represent a spectrally degenerate spatiotemporal field in which all the temporal frequencies $\omega$ lie along the \textit{same} circle in the Fourier plane [Fig.~\ref{fig:XwaveCones}(g)], although each $\omega$ is associated with a different radial frequency [Fig.~\ref{fig:XwaveCones}(f)]. This spectral degeneracy facilitated the optical realization of X-waves using conventional Bessel-beam synthesizers \cite{Durnin87PRL,Mazilu10LPR}. The spatiotemporal intensity profile $I(r,z;t)\!=\!|E(r,z;t)|^{2}$ is X-shaped in any meridional plane that includes the $z$-axis [Fig.~\ref{fig:XwaveCones}(h)].

\section{Case III: Space-time wave packets}

The hallmark of the propagation-invariant pulsed beams known as STWPs is a linear constraint involving $k_{z}$ and $\omega$ in free space:
\begin{equation}
\Omega=(k_{z}-k_{\mathrm{o}})\widetilde{v},
\end{equation}
where $\widetilde{v}$ is the group velocity, which can take on arbitrary values: subluminal ($\widetilde{v}\!<\!c$) or superluminal ($\widetilde{v}\!>\!c$) \cite{Yessenov19PRA}. The spectral of an STWP on the light-cone surface is its intersection with a plane that is parallel to the $k_{r}$-axis and makes an angle $\theta$ with the $k_{z}$-axis, so that $\widetilde{v}\!=\!c\tan{\theta}=c/\widetilde{n}$ ($\widetilde{n}\!=\!\cot{\theta}$ is the group index). Unlike X-waves, this plane does not pass through the origin, and instead passes through the point $(k_{r},k_{z},\tfrac{\omega}{c})\!=\!(0,k_{\mathrm{o}},k_{\mathrm{o}})$. The field $E(r,z;t)\!=\!e^{i(k_{\mathrm{o}}z-\omega_{\mathrm{o}}t)}\psi(r,z;t)$ has the propagation-invariant envelope:
\begin{equation}
\psi(r,z;t)=\int\!k_{r}dk_{r}\,\widetilde{\psi}(k_{r})J_{0}(k_{r}r)e^{-i\Omega(t-z/\widetilde{v})}=\psi(r,0;t-z/\widetilde{v}).
\end{equation}
The relationship between $k_{r}$ and $\omega$ in the paraxial domain is:
\begin{equation}\label{eq:STWPRadialSpatialFrequency}
\frac{\Omega}{\omega_{\mathrm{o}}}=\frac{k_{r}^{2}}{2k_{\mathrm{o}}^{2}(1-\widetilde{n})}.
\end{equation}
Here $\omega_{\mathrm{o}}$ is not the central frequency in the spectrum, but is either the maximum or minimum frequency, so that $\Omega$ is either negative- or positive-valued, which is then associated with either subluminal or superluminal STWPs, respectively.

\begin{figure}[t!]
    \centering
    \includegraphics[width=8.6cm]{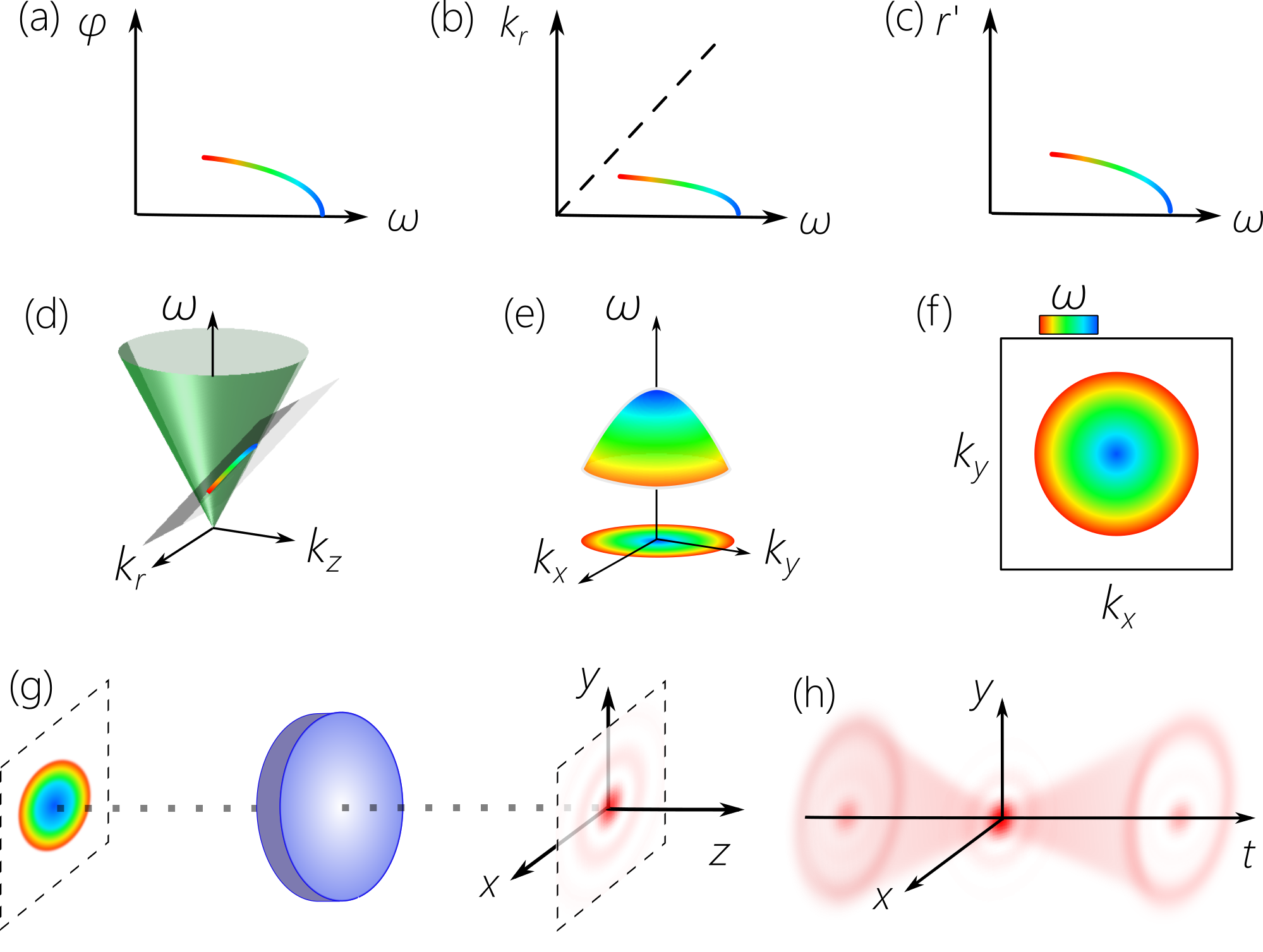}
    \caption{Spatiotemporal structure of a subluminal STWP (following Fig.~\ref{fig:PulsedBessel}). (a) The spectral AD profile $\varphi(\omega)$. (b) The spatiotemporal spectrum $\widetilde{\psi}(k_{r},\omega)$ is a segment of an ellipse. (c) The spectral distribution in the Fourier plane. (d) The spectral support is an ellipse at the intersection of the light-cone with a tilted plane that makes an angle $\theta$ with the $k_{z}$-axis. (e) The spatiotemporal spectrum in $(k_{x},k_{y},\tfrac{\omega}{c})$-space is an ellipsoid of revolution. (f) Projection of the spectrum in (e) onto the $(k_{x},k_{y})$-plane. (g) The field distribution in the back and front focal planes of a $2f$ lens. (h) In physical space, the cylindrically symmetric STWP is X-shaped in a meridional plane.}
    \label{fig:SubluminalCones}
\end{figure}

\subsection{Subluminal STWPs}

Here $\Omega\!<\!0$ and $\widetilde{n}\!>\!1$ (Eq.~\ref{eq:STWPRadialSpatialFrequency}), corresponding to $0\!<\!\theta\!<\!45^{\circ}$. Expressing $k_{r}(\omega)\!=\!\tfrac{\omega}{c}\sin\{\varphi(\omega)\}$, the propagation angle is:
\begin{equation}\label{eq:ADforSTWP}
\sin\{\varphi(\Omega)\}=\frac{\omega_{\mathrm{o}}}{\omega}\sqrt{2\frac{\Omega}{\omega_{\mathrm{o}}}(1-\widetilde{n})}.
\end{equation}
The spectral profile of $\varphi(\omega)$ is plotted in Fig.~\ref{fig:SubluminalCones}(a). When $\omega\!\rightarrow\!\omega_{\mathrm{o}}$, $\Omega\!\rightarrow\!0$, $\sin\varphi\!\rightarrow\varphi$, and we have $\varphi(\Omega)\!\propto\!\sqrt{|\Omega|}$. Because $\sqrt{x}$ is not differentiable at $x\!=\!0$, the conical-AD associated with this STWP is said to be non-differentiable \cite{Hall21OL,Hall22OEConsequences,Hall22JOSAA}; that is, $\tfrac{d\varphi}{d\omega}$ does \textit{not} exist at $\omega\!=\!\omega_{\mathrm{o}}$. Although no single optical component produces such an AD profile, a spatiotemporal Fourier synthesizer can (see next Section). The distribution of the radial spatial frequency is plotted in Fig.~\ref{fig:SubluminalCones}(b), and the radial distribution of temporal frequencies is plotted in Fig.~\ref{fig:SubluminalCones}(c). The radial position of the spatiotemporal spectrum in the Fourier plane can extend in principle to the origin $r\!=\!0$, which is associated with the maximum temporal frequency $\omega_{\mathrm{o}}$, and lower temporal frequencies (longer wavelengths) are located at larger radii.

\begin{figure}[t!]
    \centering
    \includegraphics[width=8.6cm]{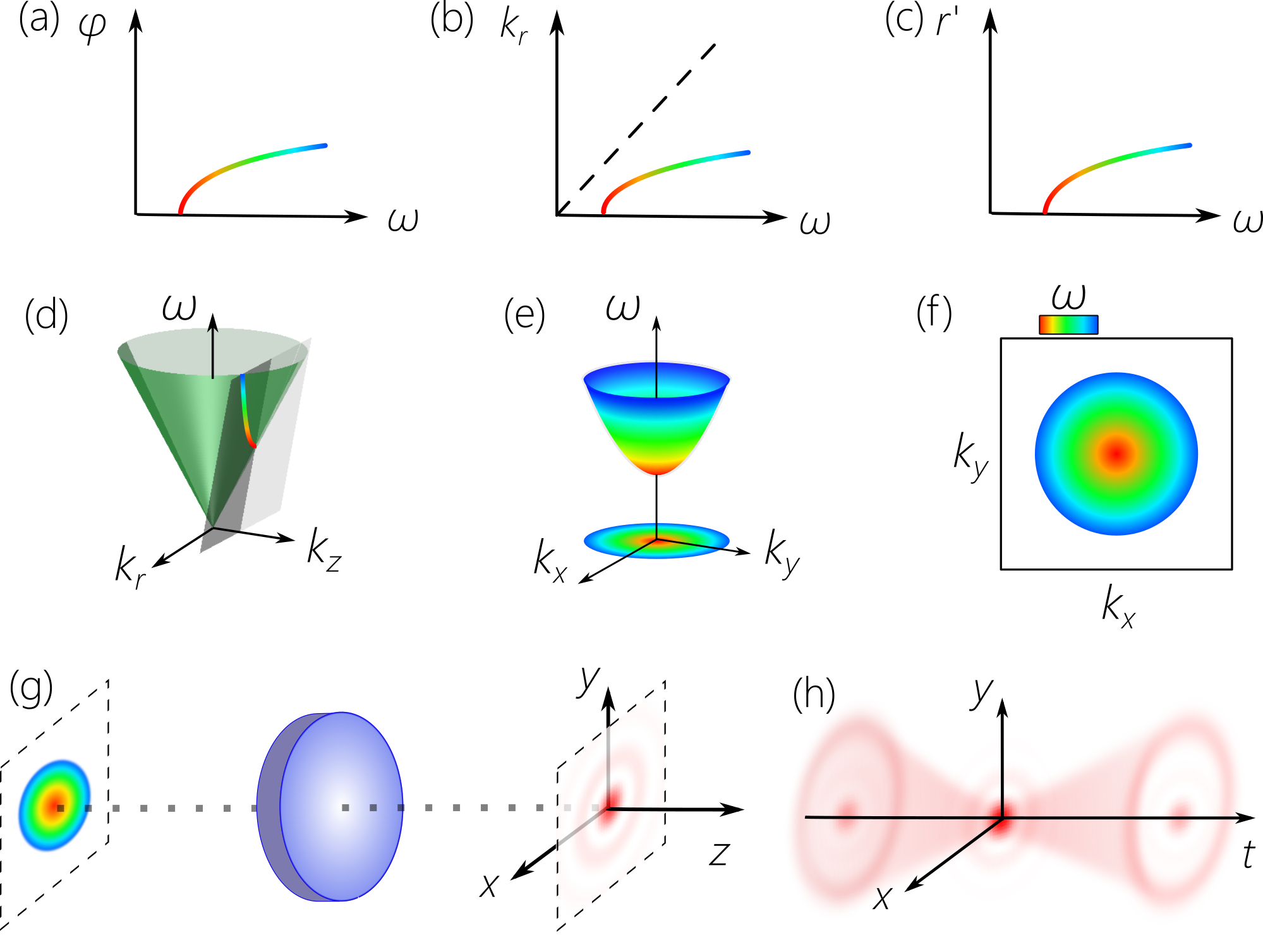}
    \caption{Spatiotemporal structure of a superluminal STWP (following Fig.~\ref{fig:PulsedBessel}). (a) The spectral AD profile $\varphi(\omega)$. (b) The spatiotemporal spectrum $\widetilde{\psi}(k_{r},\omega)$ is a segment of a hyperbola. (c) The spectral distribution in the Fourier plane. (d) The spectrum is a hyperbola at the intersection of the light-cone with a tilted plane making an angle $\theta$ with the $k_{z}$-axis. (e) The spatiotemporal spectrum in $(k_{x},k_{y},\tfrac{\omega}{c})$-space is a hyperboloid of revolution. (f) Projection of the spectrum in (e) onto the $(k_{x},k_{y})$-plane. (g) The field distribution in the back and front focal planes of a $2f$ lens. (h) In physical space, the cylindrically symmetric STWP is X-shaped in a meridional plane.}
    \label{fig:SuperluminalCones}
\end{figure}

\begin{figure*}[t!]
    \centering
    \includegraphics[width=16cm]{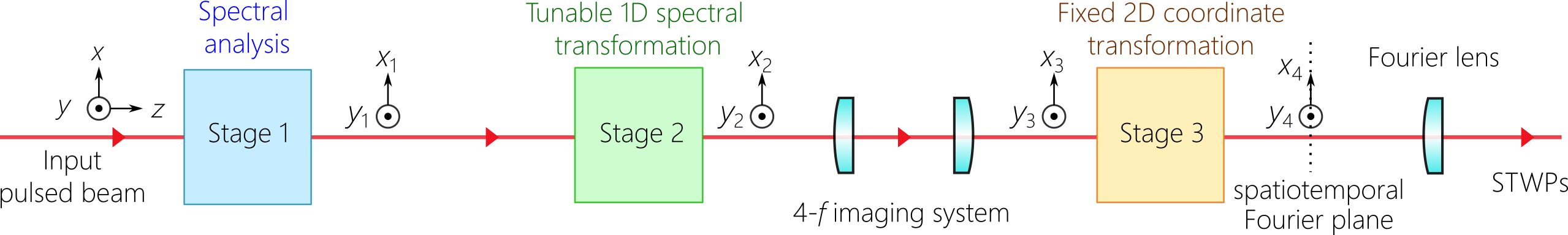}
    \caption{The synthesis of STWPs localized in all dimensions via spatiotemporal Fourier synthesis starting with a plane-wave pulse. The spectral analysis stage introduces linear spatial chirp in the $(x_{1},y_{1})$-plane along $x_{1}$. The tunable 1D spectral transformation prepares the wavelengths in a prescribed order in the $(x_{2},y_{2})$-plane, which is then imaged to the $(x_{3},y_{3})$-plane. Finally, a fixed 2D log-polar coordinate transformation converts the linearly arranged wavelengths into a radial distribution of wavelengths in the Fourier plane $(x_{4},y_{4})$. A spherical lens implements the Fourier transform.}
    \label{fig:OverallSetup}
\end{figure*}

The spectral support is a section of an ellipse at the intersection of the light-cone with a plane making an angle $\theta$ with the $k_{z}$-axis [Fig.~\ref{fig:SubluminalCones}(d)]. In $(k_{x},k_{y},\tfrac{\omega}{c})$-space, the spatiotemporal spectrum in the Fourier plane is a disk, where the highest frequency (shortest wavelength) is located at the center [Fig.~\ref{fig:SubluminalCones}(e,f)]. The spatiotemporal intensity profile of the subluminal STWP resembles that of an X-wave, except for the lower numerical aperture and smaller temporal bandwidths needed [Fig.~\ref{fig:SubluminalCones}(g,h)].

\subsection{Superluminal and negative-group-velocity STWPs}

Here $\Omega\!>\!0$ and $\widetilde{n}\!<\!1$ (Eq.~\ref{eq:STWPRadialSpatialFrequency}), corresponding to $45^{\circ}\!<\!\theta\!<\!90^{\circ}$. When $\theta\!>\!90^{\circ}$, the group velocity is negative $\widetilde{v}\!<\!0$. The propagation angle is here given by Eq.~\ref{eq:ADforSTWP}, which again corresponds to non-differentiable conical-AD in the vicinity of $\omega\!=\!\omega_{\mathrm{o}}$.

The spectral distribution of $\varphi(\omega)$ is plotted in Fig.~\ref{fig:SuperluminalCones}(a), that of the radial spatial frequency in Fig.~\ref{fig:SuperluminalCones}(b), and the radial distribution of temporal frequencies $r(\omega)$ in Fig.~\ref{fig:SuperluminalCones}(c). The spectral support is the hyperbola at the intersection of the light-cone with the tilted plane when $45^{\circ}\!<\!\theta\!<\!135^{\circ}$ [Fig.~\ref{fig:SuperluminalCones}(d)], a parabola when $\theta\!=\!135^{\circ}$, and an ellipse when $135^{\circ}\!<\!\theta\!<\!180^{\circ}$.  In contrast to subluminal STWPs, the lowest temporal frequency $\omega_{\mathrm{o}}$ is located at the origin of the Fourier plane [Fig.~\ref{fig:SuperluminalCones}(e,f)]. The spatiotemporal intensity profile resembles that of a subluminal STWP [Fig.~\ref{fig:SuperluminalCones}(g,h)].

\section{Experimental realization}

Synthesizing STWPs that are localized in all dimensions via spatiotemporal Fourier synthesis poses a considerable challenge by requiring a particular arrangement of wavelengths in a prescribed sequence of precisely defined annuli [Fig.~\ref{fig:SpatiotemporalSpectrum}]. This challenge went unmet in early research on localized waves \cite{Yessenov22AOP}, whether FWMs, pulsed Bessel beams, or other theoretically identified propagation-invariant wave packets in free space or in dispersive media \cite{Malaguti08OL,Malaguti09PRA,Turunen10PO,FigueroaBook14}. The simplicity of X-waves motivated their study, even though they are virtually a dead-end in optics as elucidated in \cite{Yessenov22AOP}. Early work on preparing STWPs in the form of light-sheets (localized along one transverse dimension and uniform along the other) \cite{Kondakci17NP,Kondakci18OE,Yessenov19OE,Yessenov19OPN} has been recently generalized to a spatiotemporal Fourier synthesizer \cite{Yessenov22NC,Yessenov22OL,Yessenov25Meron}, which we present her. We describe briefly in the Discussion other recently emerging approaches.

\begin{figure}[t!]
    \centering
    \includegraphics[width=8.6cm]{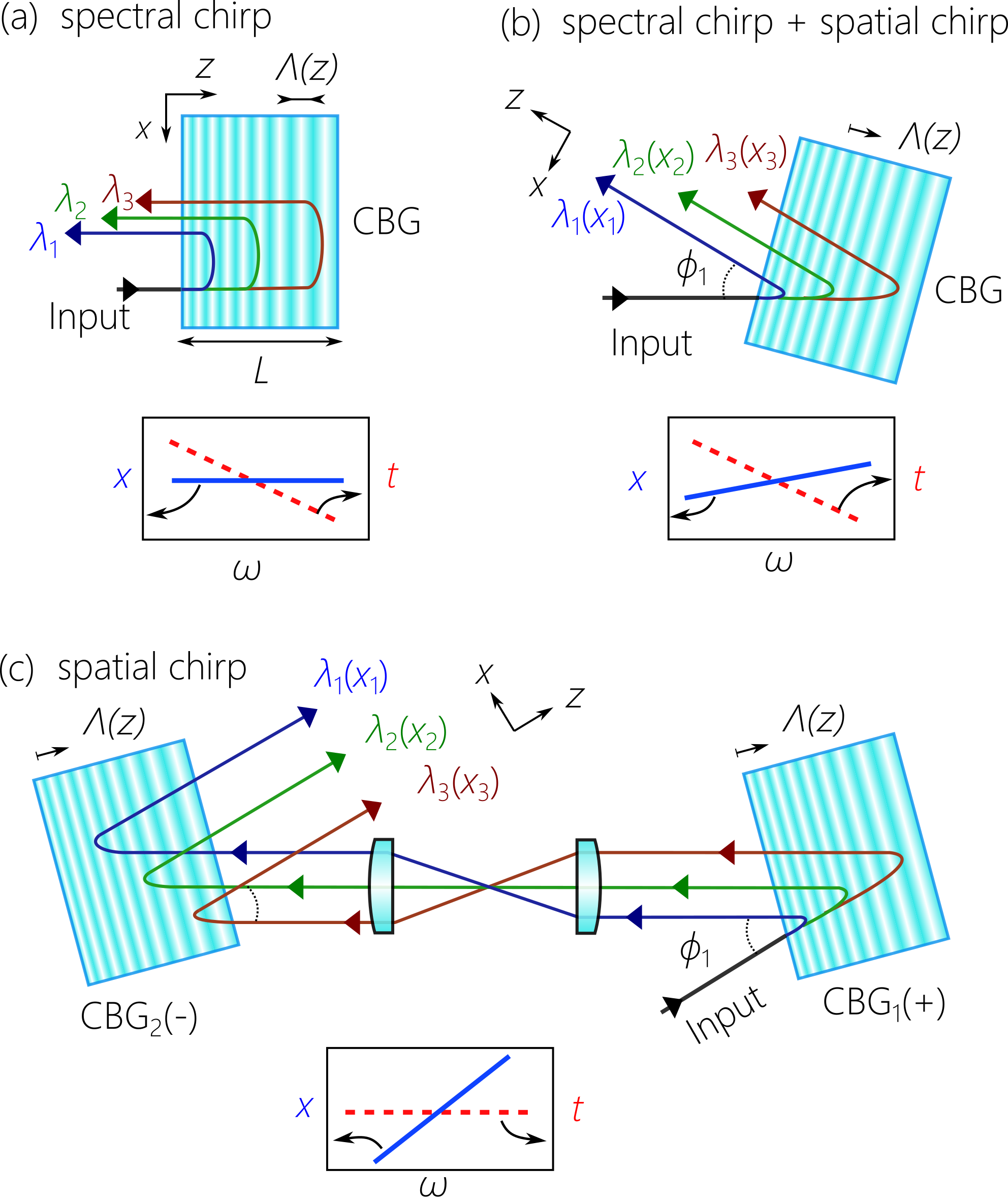}
    \caption{Spectral analysis via CBGs. (a) At normal incidence, wavelengths reflect from different depths $z$, thus introducing spectral chirp, but no \textit{spatial} chirp. The bottom panel depicts the transverse position $x$ of the temporal frequencies $\omega$ (the spectrum is \textit{not} spatially resolved) and the group delay. (b) At oblique incidence, spatial and spectral chirps are produced. The bottom panel depicts the corresponding quantities plotted in the bottom panel in (a). (c) In a double-pass configuration using two identical CBGs having oppositely signed chirps, the second pass eliminates the spectral chirp from the first pass and doubles the spatial chirp. The $\times2$ de-magnification in the $4f$ relay from the first CBG to the second pre-compensates for the spatial-chirp doubling. The bottom panel depicts the linear spatial chirp and zero spectral chirp.}
    \label{fig:CBGconcept}
\end{figure}

\subsection{Overall strategy}

The overall strategy for spatiotemporal Fourier synthesis to produce STWPs localized in all dimensions starting from a generic pulsed laser is depicted in Fig.~\ref{fig:OverallSetup}. Three stages can be identified. In the first stage, the spectrum of the initial pulse is spatially resolved (stage~1: spectral analysis). The wavelengths are arranged in space in a fixed sequence along one dimension. In the second stage, the wavelengths are re-arranged spatially to realize a prescribed order via a 1D spectral transformation (stage~2: spectral re-organization). The third stage then converts this sequence of wavelengths arranged along one dimension in Cartesian coordinates to an azimuthally symmetric radial sequence of wavelengths via a 2D log-polar coordinate transformation (stage~3: fixed 2D coordinate transformation). The output of this third stage constitutes the spatiotemporal Fourier plane, and a spherical lens converts this spatiotemporal spectrum into physical space. We proceed to describe each of these stages.

\subsection{Stage~1: Spectral analysis}

To spatially resolve the spectrum of the initial pulse, we make use of chirped volume Bragg gratings (CBGs) \cite{Kaim14OEng,Glebov14OEng} rather than a conventional diffraction grating because the subsequent spectral and spatial transformation stages require a collimated wave front. At normal incidence, the wavelengths reflect from different depths, resulting in spectral chirp and pulse stretching [Fig.~\ref{fig:CBGconcept}(a)]. At oblique incidence, spectral chirp is combined with \textit{spatial} chirp: each wavelength is spatially displaced in the transverse plane in proportion to the depth from which it reflects [Fig.~\ref{fig:CBGconcept}(b)]. Combining two CBGs in the configuration shown in Fig.~\ref{fig:CBGconcept}(c) eliminates the spectral chirp and doubles the spatial chirp. The result is a collimated spatially resolved spectrum.

\begin{figure}[t!]
    \centering
    \includegraphics[width=8.6cm]{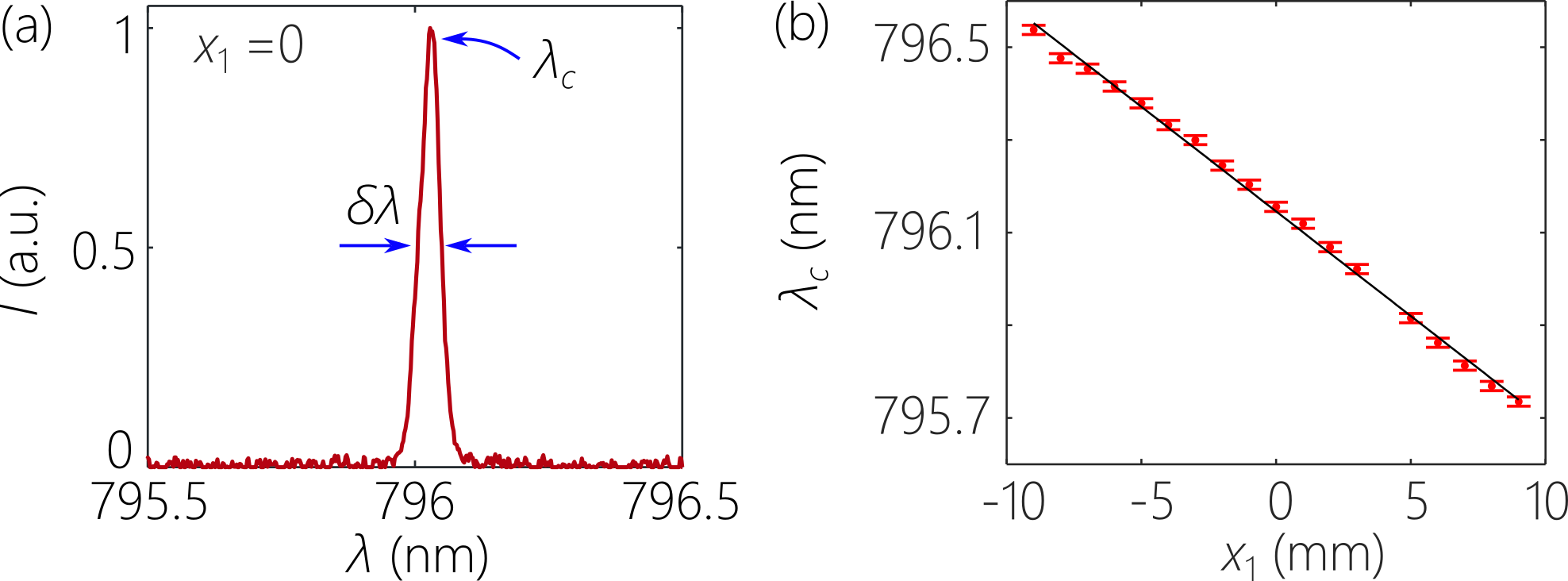}
    \caption{(a) The measured spectrum at $x_{1}\!=\!0$ (Fig.~\ref{fig:OverallSetup} after stage~1) collected by a single-mode fiber. (b) The measured spatial spread of the spectrum collected by repeating the measurement in (a) over $x_1$. The straight line is a theoretical fit, the symbols are data, and the error bars correspond to the spectral resolution $\approx10$~pm of the optical spectrum analyzer used.}
    \label{fig:CBGdata}
\end{figure}

If the CBG has a linearly varying periodicity $\Lambda(z)$, and the pulses are incident at an angle $\phi_{1}$ with the normal to the CBG input facet, then each wavelength $\lambda$ emerges at a position:
\begin{equation}
x(\lambda)=\frac{1}{\beta}(\lambda-\lambda_{\mathrm{o}})\xi(\phi_{1}),
\end{equation}
where $\xi(\phi_{1})\!=\!\tfrac{n\sin2\phi_{1}}{n^{2}-\sin^{2}\phi_{1}}$, $\lambda_{\mathrm{o}}\!=\!2n\Lambda_{\mathrm{o}}$ is the central wavelength, $\Lambda_{\mathrm{o}}$ is the Bragg period at the CBG center, $n$ is the average index of the CBG, and $\beta\!=\!\tfrac{d\lambda}{dz}\!=\!2n\tfrac{d\Lambda}{dz}$ is the CBG chirp rate.

Starting with 100-fs pulses from a mode-locked Ti:sapphire laser (Tsunami, Spectra Physics) of bandwidth $\Delta\lambda\!\approx\!10$~nm at a wavelength $\approx\!800$~nm, we direct the field to a CBG (Optigrate L1-021) having $\Lambda_{\mathrm{o}}\!=\!270$~nm, $n\!=\!1.5$, length $L\!=\!34$~mm, chirp rate $\beta\!=\!-30$~pm/mm, and incident angle $\phi_{1}\!=\!16^{\circ}$. After reflection from the first CBG, the field is characterized by both spatial and spectral chirp. We relay this field to a second identical CBG with reversed chirp (we make use of the opposite port of the same CBG) via a $4f$ imaging system that de-magnifies the transverse width by a factor of 2 to undue the additional spatial chirp. The result is a collimated, spectral-chirp-free spatially resolved spectrum along $x_{1}$ in the $(x_{1},y_{1})$-plane:
\begin{equation}
x_{1}(\lambda)=\alpha(\lambda-\lambda_{1}),
\end{equation}
where $\alpha\!=\!\tfrac{1}{\beta}\xi(\phi_{1})$ is the spatial chirp rate. In our experiment we have $\lambda_{\mathrm{o}}\!=\!796.1$~nm and $\alpha\!\approx\!22.2$~mm/nm. From the measured spatially resolved spectrum [Fig.~\ref{fig:CBGdata}], we identify the spectral uncertainty produced by the CBG [Fig.~\ref{fig:CBGdata}(a)] and the linear variation of wavelength with transverse position [Fig.~\ref{fig:CBGdata}(b)]. 

\subsection{Stage~2: Spectral reorganization}

The spatially resolved spectrum has a fixed arrangement of wavelengths in the $(x_{1},y_{1})$-plane (Fig.~\ref{fig:OverallSetup}; stage~1). To produce a prescribed spatiotemporal spectrum in the Fourier plane, we need to: (1) reorganize the sequence of wavelengths, \textit{and} (2) change the geometry from Cartesian to a polar coordinate system. Although it is conceivable that a single optical system could perform both of these tasks, we choose to perform them separately for simplicity (see below).

\begin{figure}[t!]
    \centering
    \includegraphics[width=8.6cm]{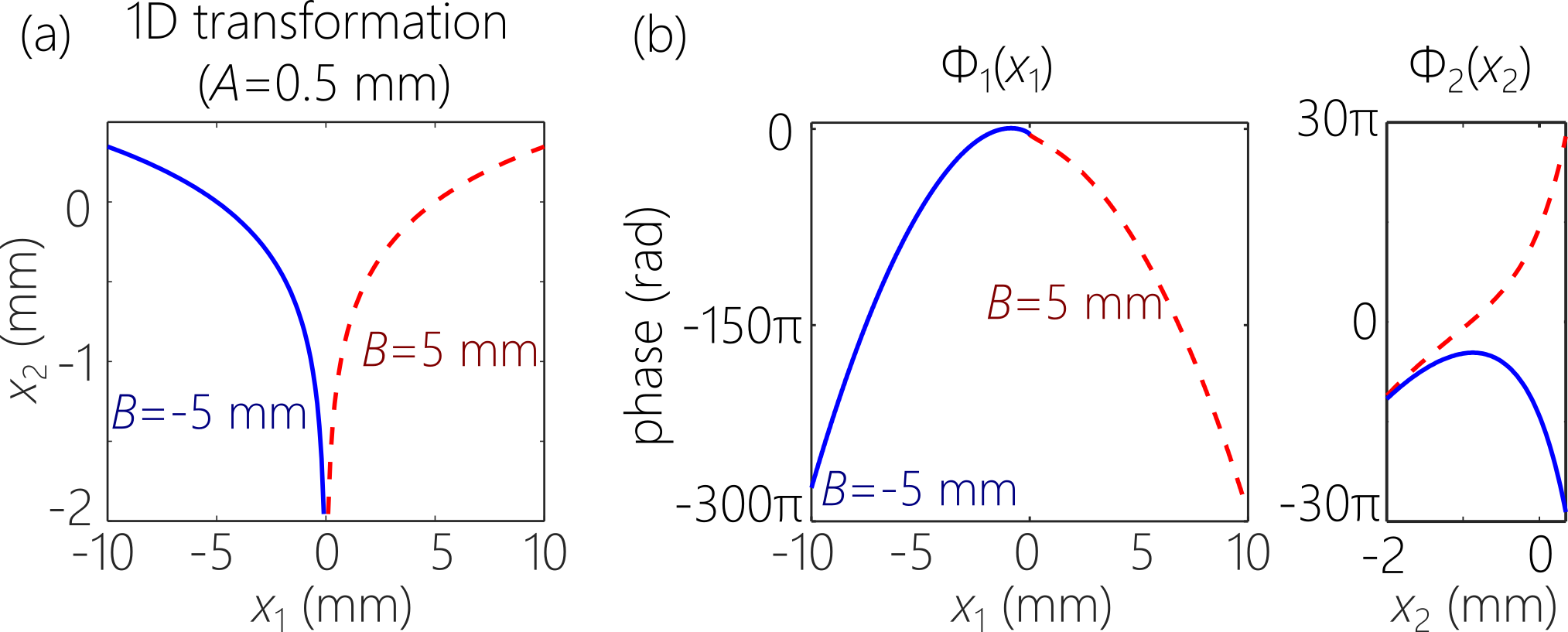}
    \caption{Spectral reorganization via a 1D spatial coordinate transformation. (a) The transformation between the coordinate systems $(x_{1},y_{1})$ and $(x_{2},y_{2})$, with $y_{2}\!=\!y_{1}$. We have $(A,B)\!=\!(0.5,-5)$~mm (solid curve) and $(A,B)\!=\!(0.5,5)$~mm (dashed curve). (b) Plots of the phase distributions $\Phi_{1}(x_{1})$ and $\Phi_{2}(x_{2})$ to implement the transformations in (a).}
    \label{fig:1DTransform}
\end{figure}

Spectral reorganization is performed by an optical system that maps the $(x_{1},y_{1})$-plane to the $(x_{2},y_{2})$-plane according to:
\begin{equation}\label{eq:1Dmapping}
x_{2}=A\mathrm{ln}\left(\frac{x_{1}}{B}\right),\;\;\;y_{2}=y_{1},
\end{equation}
where $A$ and $B$ are the transformation parameters, and the field remains uniform along $y$. Because the wavelengths are arranged linearly along $x_{1}$ at the input, the spectrum is reorganized along $x_{2}$ at the output. The parameter $A$ is a scaling parameter along $x_{2}$, while the parameter $B$ scales the coordinate $x_{1}$ before taking the logarithm. Because the argument to the logarithm must be positive, selecting positive or negative values for $B$ requires changing the valid domain of $x_{1}$: the positive $x_{1}$-axis when $B\!>\!0$ and the negative $x_{1}$-axis when $B\!<\!0$; see Fig.~\ref{fig:1DTransform}. The logarithm introduced in this transformation will be undone in the subsequent 2D log-polar coordinate transformation that follows this stage.

We implement the coordinate transformation in Eq.~\ref{eq:1Dmapping} using a pair of phase distributions $\Phi_{1}(x_{1},y_{1})$ and $\Phi_{2}(x_{2},y_{2})$ placed at the $(x_{1},y_{1})$ and $(x_{2},y_{2})$ planes, respectively, having the form:
\begin{eqnarray}
\Phi_{1}(x_{1},y_{1})&=&\frac{kA}{d_{1}}
\left\{x_{1}\mathrm{ln}\left(\frac{x_{1}}{B}\right)-x_{1}\right\}-\frac{kx_{1}^{2}}{2d_{1}},\nonumber\\
\Phi_{2}(x_{2},y_{2})&=&\frac{kAB}{d_{1}}\exp\left(\frac{x_{2}}{A}\right)-\frac{kx_{2}^{2}}{2d_{1}},
\end{eqnarray}
where $k\!=\!\tfrac{\omega}{c}$ is the wave number. Instead of placing a lens between the two phase distributions, we add the quadratic phase terms $\tfrac{kx_{1}^{2}}{2d_{1}}$ and $\tfrac{kx_{2}^{2}}{2d_{1}}$ to implement a lens of focal length $d_{1}$ \cite{Yessenov22NC}.

The two phase distributions $\Phi_{1}(x_{1},y_{1})$ and $\Phi_{2}(x_{2},y_{2})$ are implemented using a pair of SLMs (Meadowlark, $1920\times1080$ series), with $d_{1}\!=\!400$~mm, fixing the parameter $A\!=\!0.5$~mm, and varying the second parameter $B$ in the range $B\!=\![-15,20]$~mm. The two SLMs are followed by a $4f$ imaging system comprising two lenses to relay the field from the $(x_{2},y_{2})$-plane to the $(x_{3},y_{3})$-plane at the entrance to the 2D log-polar coordinate transformation, where $x_{3}\!=\!-x_{2}$ and $y_{3}\!=\!y_{2}$. We place a beam stop in the Fourier plane of the $4f$ system to eliminate the zeroth-order (non-scattered) field component from the two SLMs.

\subsection{Stage~3: 2D Log-polar coordinate transformation}

\begin{figure}[t!]
    \centering
    \includegraphics[width=8.6cm]{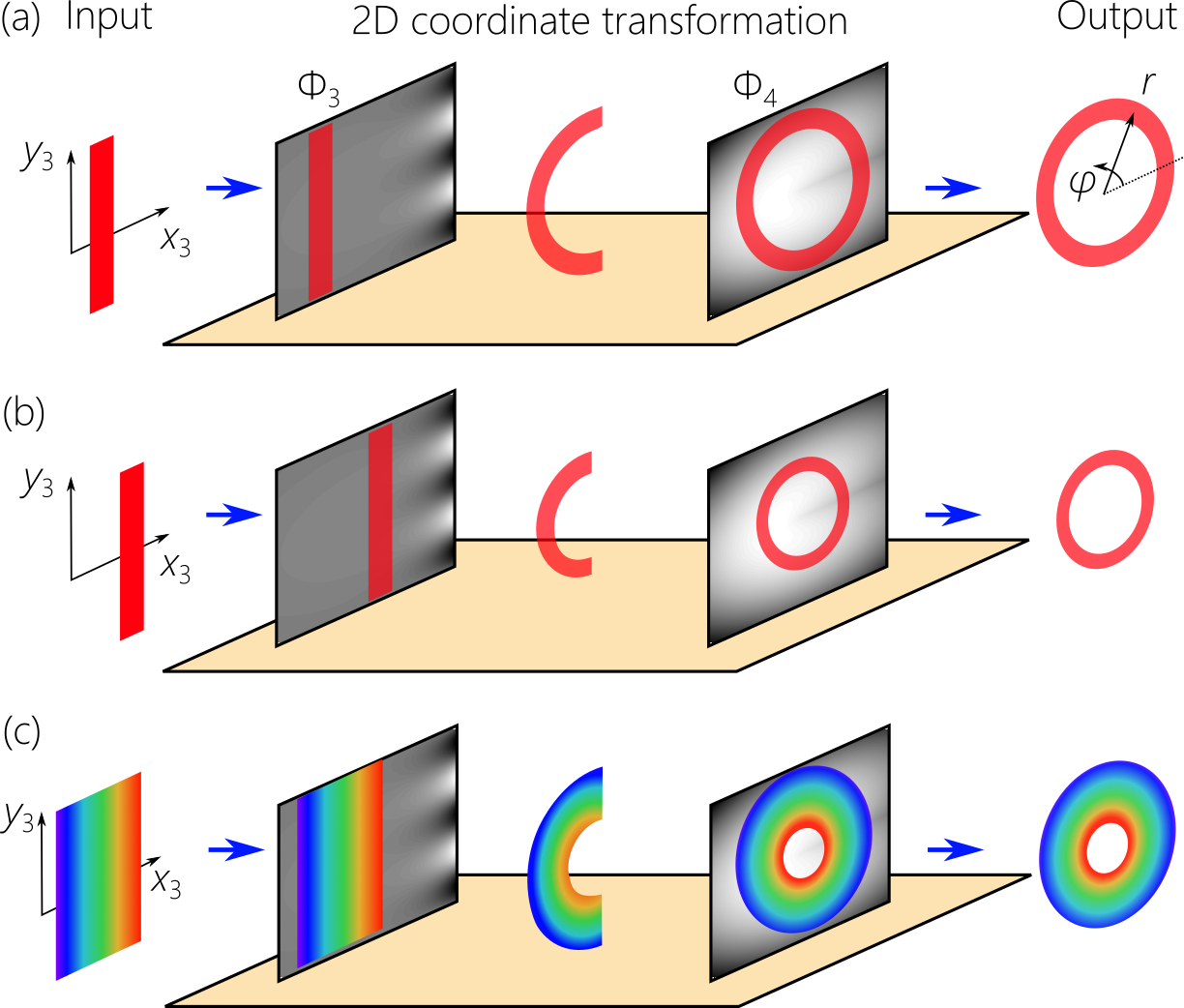}
    \caption{Preparing the spatiotemporal spectrum in the Fourier plane via a log-polar coordinate transformation. (a) A monochromatic rectangular strip of light gradually curves after $\Phi_{3}$ and becomes a closed annulus at the $(x_{4},y_{4})$-plane, where $\Phi_{4}$ is a correction phase. (b) Displacing the strip in (a) along $x_{3}$ changes the radius of the resulting annulus. (c) When the input is a spectrally resolved wave front along $x_{3}$, each wavelength is converted into an annulus of a different radius to form the spatiotemporal spectrum in the Fourier plane.}
    \label{fig:2DTransformConcept}
\end{figure}

In the $(x_{3},y_{3})$-plane, the wavelengths are arranged in a Cartesian coordinate system along $x_{3}$ (the field is uniform along $y_{3}$). We aim at transforming the coordinate system to arrange the wavelengths radially along azimuthally symmetric annuli in the $(x_{4},y_{4})$-plane; see Fig.~\ref{fig:2DTransformConcept}. This corresponds to the following 2D coordinate transformation:
\begin{equation}
r=C\exp\left(-\frac{x_{3}}{D}\right),\;\;\;\phi=\frac{y_{3}}{D},
\end{equation}
where $r\!=\!\sqrt{x_{4}^{2}+y_{4}^{2}}$, $\phi\!=\!\mathrm{arctan}(\tfrac{y_{4}}{x_{4}})$, $D$ is a parameter that maps the vertical size of the input $y_{3}\!=\![-y_{3}^{\mathrm{max}},y_{3}^{\mathrm{max}}]$ to the angular span $\phi\!=\![-\pi,\pi]$, $D\!=\!y_{3}^{\mathrm{max}}/\pi$, and $C$ is limited by the aperture.  

\begin{figure}[t!]
    \centering
    \includegraphics[width=8.6cm]{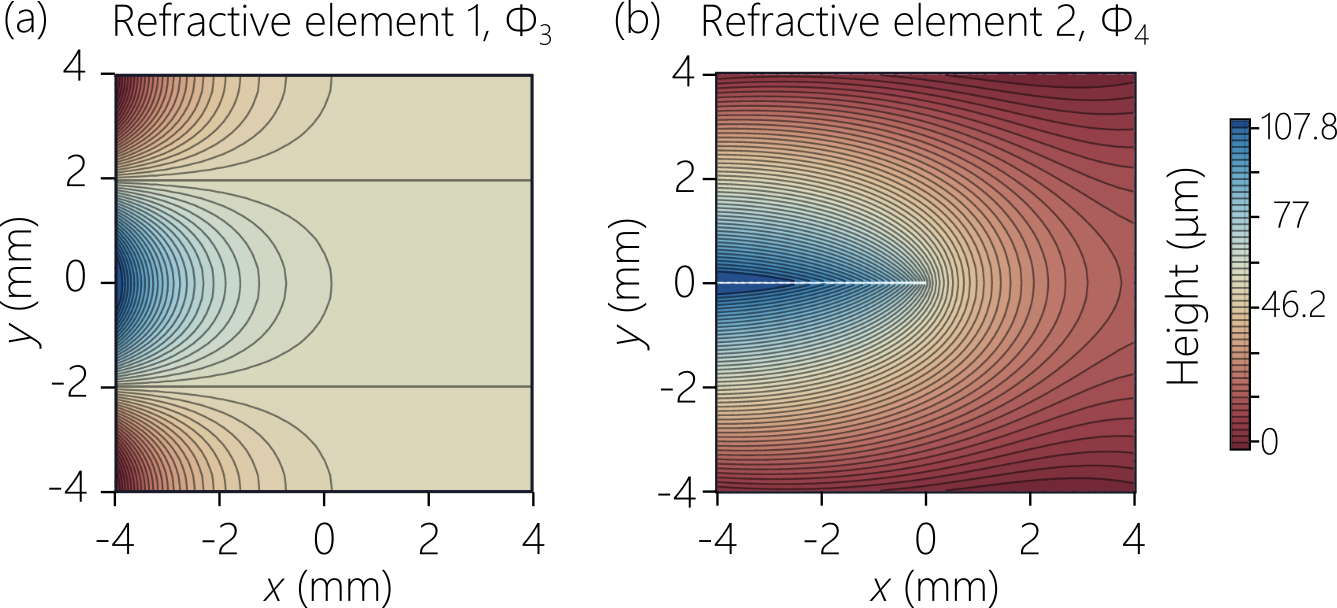}
    \caption{(a,b) Height profiles of refractive phase plates for implementing the 2D coordinate transformation. The phase plates have an aperture size of $2y_{3}^{\mathrm{max}}=8$~mm, separated by $d_{2}=310$~mm, with $C=4.77$~mm and $D=1$~mm. }
    \label{fig:2DPhasePlate}
\end{figure}

This coordinate transformation is implemented by two phase distributions $\Phi_{3}(x_{3},y_{3})$ and $\Phi_{4}(x_{4},y_{4})$ in the $(x_{3},y_{3})$ and $(x_{4},y_{4})$ planes, respectively, which are given by:
\begin{eqnarray}
\Phi_{3}(x_{3},y_{3})\!\!\!\!&=&\!\!\!\!\!-\frac{kCD}{d_{2}}\exp\left(-\frac{x_{3}}{D}\right)\cos\left(\frac{y_{3}}{D}\right)-\frac{k}{2d_{2}}(x_{3}^{2}+y_{3}^{2}),\nonumber\\
\Phi_{4}(x_{4},y_{4})\!\!\!\!&=&\!\!\!\!\!\frac{kD}{d_{2}}\left\{\mathrm{atan2}(y_{4},x_{4})-x_{4}\mathrm{ln}\left(\frac{r}{C}\right)+x_{4}\right\}-\frac{kr^{2}}{2d_{2}},
\end{eqnarray}
where $\mathrm{atan2}$ is the 2-argument arc-tangent function. In typical implementations, a lens is placed between the two phase distributions. We eliminate that lens and replace it with two appropriate phase factors of the form $\tfrac{k}{2d_{2}}(x_{3}^{2}+y_{3}^{2})$ and $\frac{kr^{2}}{2d_{2}}$ in $\Phi_{3}(x_{3},y_{3})$ and $\Phi_{4}(x_{4},y_{4})$, respectively. We plot in Fig.~\ref{fig:2DPhasePlate}(a,b) an example of these two phase distributions plotted in terms of the height of a phase plate at a wavelength $\lambda\!=\!800$~nm and refractive index $n\!\approx\!1.5$.

The phase distributions $\Phi_{3}$ and $\Phi_{4}$ can be implemented using SLMs, which have the advantage of providing changing parameters, but are restricted by their limited spatial resolution. The needed changes in this transformation (such as changing wavelength order or scaling the spectrum) are incorporated into the preceding spectral re-organization stage, so that the 2D log-polar coordinate transformation can be kept stationary. We can thus make use of implementations that provide higher efficiency, such as diamond-turned surfaces \cite{Lavery12OE} or lithographically inscribed phase plates \cite{Li19OE}.

\subsection{Overall system}

\begin{figure}[t!]
    \centering
    \includegraphics[width=8.6cm]{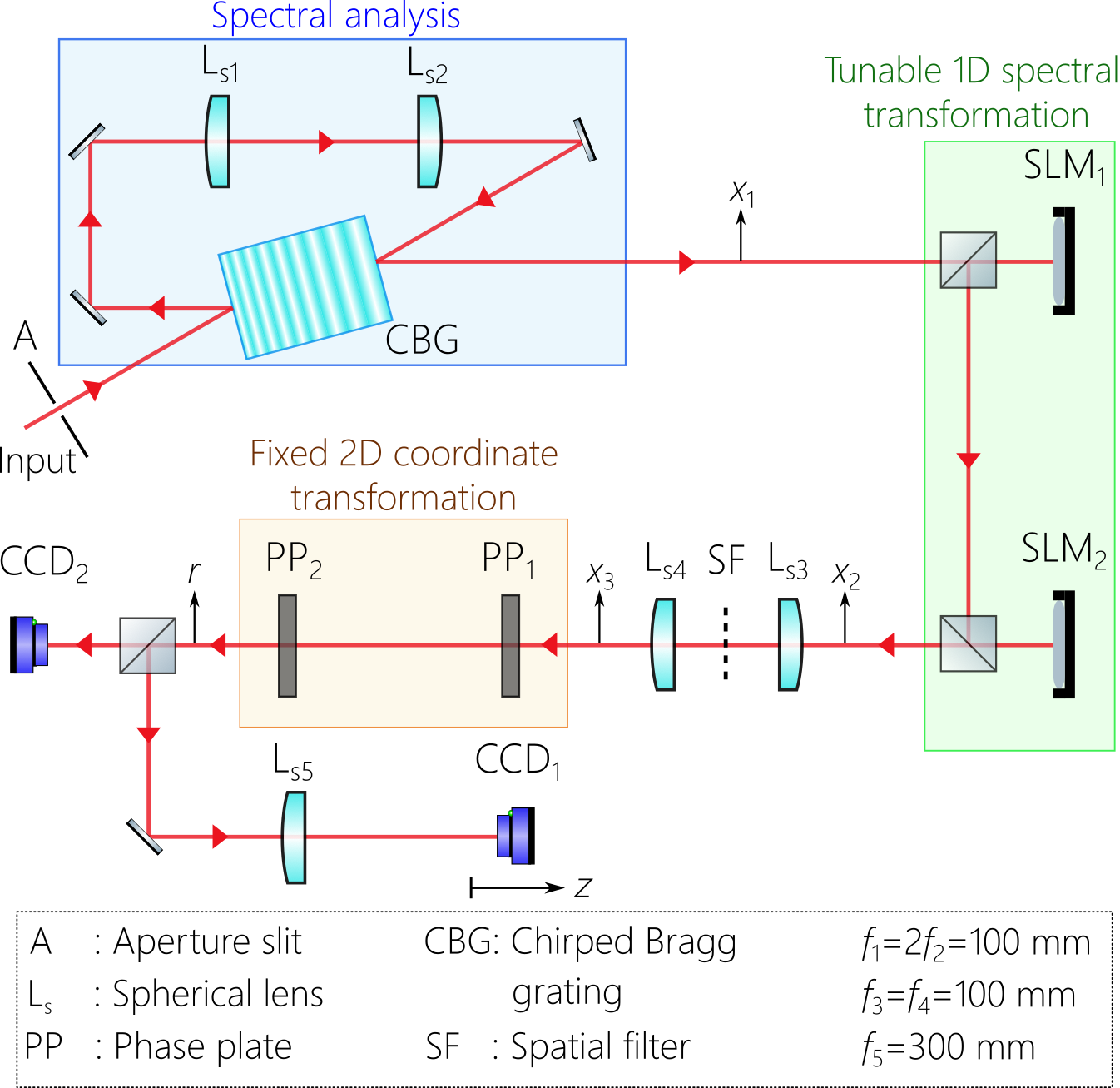}
    \caption{Overall experimental setup combining: spectral analysis, 1D spectral re-organization, 2D log-polar coordinate transformation, and the Fourier transforming lens. The two CCD cameras indicate two measurement planes: the physical plane (CCD$_{1}$) and the spatiotemporal Fourier plane (CCD$_{2}$).}
    \label{fig:OverallSystem}
\end{figure}

Starting with the linear spatial chirp in the $(x_{1},y_{1})$-plane given by $x_{1}(\lambda)\!=\!\alpha(\lambda-\lambda_{\mathrm{o}})$, the spatial chirp is modified after the 1D spectral re-organization stage, $x_{2}(\lambda)\!=\!A\mathrm{ln}\{\tfrac{\alpha}{B}(\lambda-\lambda_{\mathrm{o}})\}$. The log-normal coordinate transformation compensates for the logarithm in $x_{2}$ and produces the \textit{radial} chirp:
\begin{equation}
r(\lambda)=C\exp\left\{-\frac{x_{3}(\lambda)}{D}\right\}=C\left(\frac{\alpha}{B}[\lambda-\lambda_{\mathrm{o}}]\right)^{A/D}.
\end{equation}
By setting $D\!=\!2A$, the spatial frequency in the Fourier plane is:
\begin{equation}
k_{r}(\lambda)\approx C\frac{k_{\mathrm{o}}}{f}\sqrt{\frac{\alpha}{B}(\lambda-\lambda_{\mathrm{o}})},
\end{equation}
corresponding to Eq.~\ref{eq:STWPRadialSpatialFrequency}. In the data described next, we have $\lambda_{\mathrm{o}}\!\approx\!798$~nm, the chirp rate is $\alpha\!=\!-22.2$~mm/nm, and $C\!\approx\!4.8$~mm after limiting the aperture to $2y_{3}^{\mathrm{max}}\!=\!8$~mm. This leaves $B$ as a free parameter to tune the STWP group velocity. The overall setup corresponding to the scheme in Fig.~\ref{fig:OverallSetup} is depicted in Fig.~\ref{fig:OverallSystem}.

\begin{figure*}[t!]
    \centering
    \includegraphics[width=14cm]{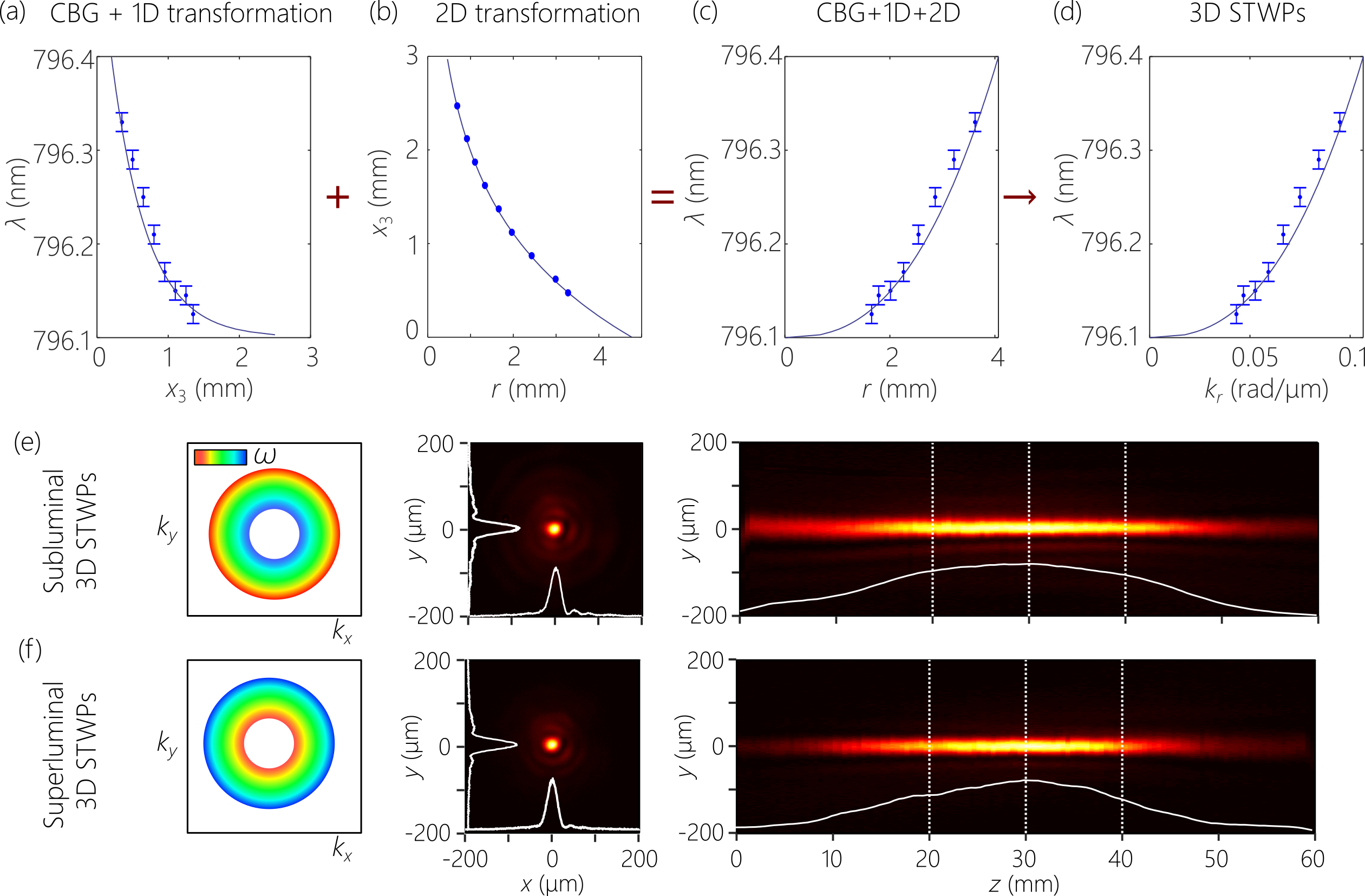}
    \caption{(a-d) Characterizing the spatiotemporal spectrum in the Fourier plane. (a) Measured spectral distribution along $x_{3}$ at the entrance to the log-polar coordinate transformation. (b) Measured mapping between the location of a slit along $x_{3}$ at the input to the log-polar transformation and the radius $r$ of the annulus produced at its output in the $(x_{4},y_{4})$-plane. (c) Determining the radius $r$ of each wavelength $\lambda$ in the Fourier plane $(x_{4},y_{4})$ by combining (a) and (b). (d) Scaling the spatiotemporal spectrum in (c) in terms of $(k_{r},\lambda)$. (e) Time-averaged intensity of the subluminal STWP with $\widetilde{v}\!=\!0.83c$ and (f) superluminal STWP with $\widetilde{v}\!=\!1.37c$. The left panel sketches the spatiotemporal spectrum. The middle panel shows the time-averaged intensity $I(x,y,z)$ at $z\!=\!30$~mm; the white curves correspond to $x\!=\!0$ and $y\!=\!0$. The right panel shows the time-averaged intensity in a meridional plane $I(0,y,z)$. The white curve is $I(0,0,z)$, and the dashed white lines are the planes where we measure the spatiotemporal intensity profiles in Fig.~\ref{fig:TimeResolvedIntensity}.}
    \label{fig:TimeAverageData}
\end{figure*}

\subsection{Characterization}

The STWPs produced by spatiotemporal Fourier synthesis are characterized in several domains to benchmark the non-separable spatiotemporal field structure. These measurements include (1) the spatiotemporal spectrum that reveals the association between the spatial and temporal frequencies; (2) the axial time-averaged intensity, which reveals the extent of its diffraction-free propagation; and (3) the spatiotemporal intensity distribution that reveals the non-separable field structure in space and time at a fixed axial plane.

\subsubsection{Spatiotemporal spectral measurement}

We measure the spatiotemporal spectrum in the Fourier plane in two steps. First, we measure the spatiotemporal spectrum in the $(x_{3},y_{3})$-plane by scanning a single-mode fiber along $x_{3}$ that is connected to an optical spectrum analyzer to determine the spatial chirp of the spectrum after spectral re-organization [Fig.~\ref{fig:TimeAverageData}(a)]. We next verify the operation of the 2D log-polar transformation by scanning a rectangular slit of light along $x_{3}$ and measure the radius $r$ of the annulus formed at the output [Fig.~\ref{fig:TimeAverageData}(b)]. Combining these two measurements, we can determine the radius of the annulus associated with each wavelength $\lambda$ in the Fourier plane [Fig.~\ref{fig:TimeAverageData}(c)], which corresponds to the synthesized spatiotemporal spectrum $|\widetilde{\psi}(k_{r},\lambda)|^{2}$ [Fig.~\ref{fig:TimeAverageData}(d)].

\subsubsection{Time-averaged intensity measurements}

We measure the time-averaged intensity distribution $I(x,y,z)\!=\!\int\!dt\,|E(x,y,z;t)|^{2}$ by scanning a CCD camera along the $z$-axis and capturing the 2D spatial intensity profile at each axial plane. We plot in Fig.~\ref{fig:TimeAverageData}(e) the circularly symmetric 2D intensity profile at the axial plane $z\!=\!30$~mm for a subluminal STWP ($\widetilde{v}\!=\!0.83c$). A cross section of the intensity distribution in a meridional plane $x\!=\!0$ passing through the $z$-axis, $I(0,y,z)$, is also plotted. The axial diffraction-free length of $\approx\!60$~mm exceeds that of a field having the same spatial bandwidth but a separable spatiotemporal spectrum, and also exceeds the propagation distance of a Gaussian beam of the same transverse spatial width. In Fig.~\ref{fig:TimeAverageData}(f), the corresponding measurements for a superluminal STWP ($\widetilde{v}\!=\!1.37c$) show similar characteristics to its subluminal counterpart. 

\subsubsection{Time-resolved intensity measurements}

\begin{figure}[t!]
    \centering
    \includegraphics[width=8.6cm]{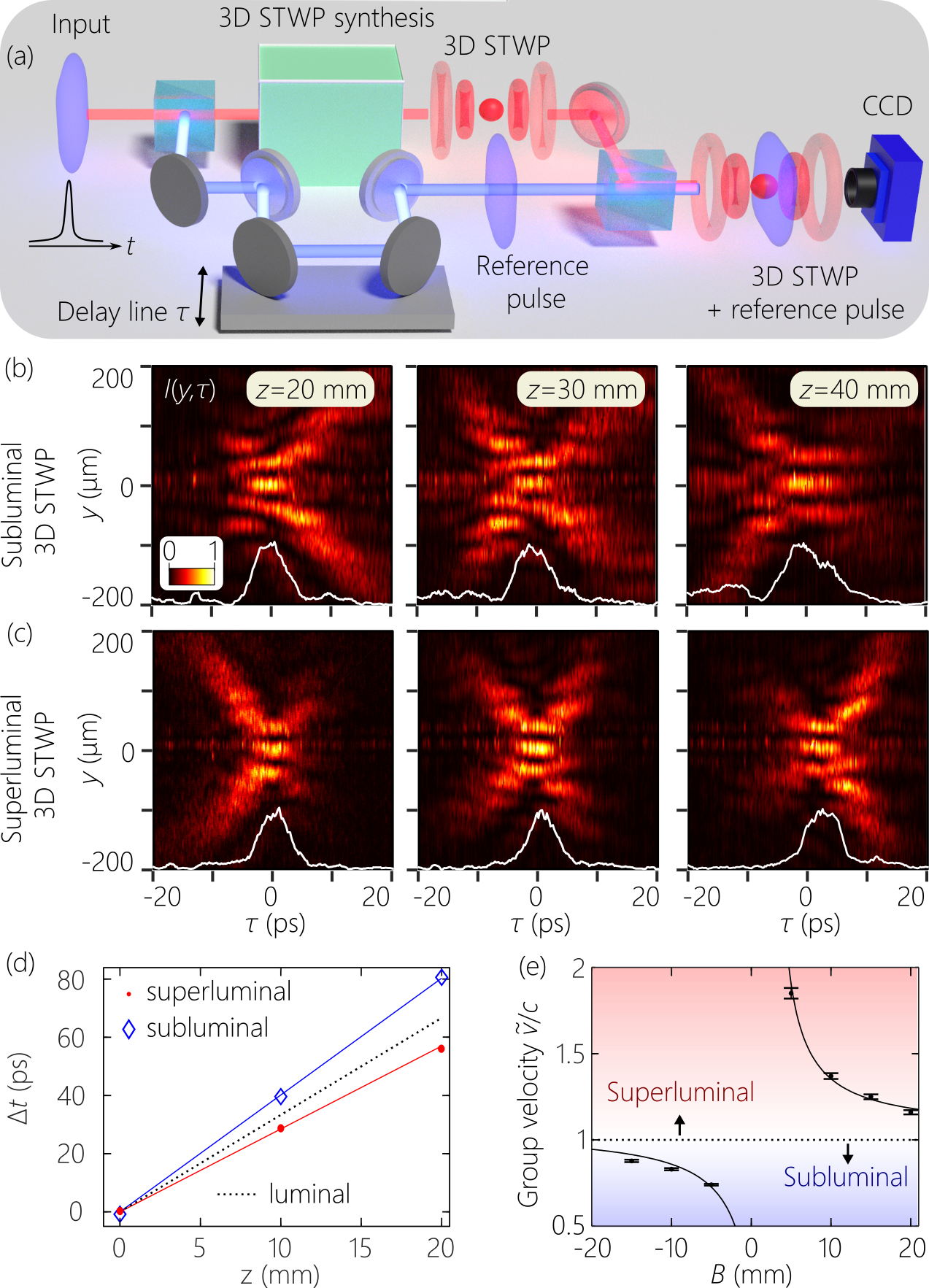}
    \caption{(a) Setup for reconstructing the spatiotemporal intensity profile $I(x,y,z;\tau)$. (b) Measured spatiotemporal intensity profiles $I(0,y,z;\tau)$ for a subluminal STWP ($\widetilde{v}\!=\!0.83c$) in the meridional plane $x\!=\!0$, obtained in three axial planes $z\!=\!20$, 30, and 40~mm, as identified in Fig.~\ref{fig:TimeAverageData}(e). (c) Same as (b) for a superluminal STWP ($\widetilde{v}\!=\!1.37c$). (d) Measured relative delay $\Delta t$ between the STWP (traveling at $\widetilde{v}$) and the reference luminal pulse (traveling at $c$) for the subluminal and superluminal STWPs in (b) and (c). (e) Measured group velocity $\widetilde{v}$ for STWPs as we vary the parameter $B$ in the spectral re-organization stage in the spatiotemporal Fourier synthesizer.}
    \label{fig:TimeResolvedIntensity}
\end{figure}

To reconstruct the spatiotemporal intensity profile $I(x,y,z;t)\!=\!|E(x,y,z;t)|^{2}$ at any axial plane $z$, we make use of the experimental configuration depicted in Fig.~\ref{fig:TimeResolvedIntensity}(a). We place the entire spatiotemporal Fourier synthesis system in one arm of a Mach-Zehnder interferometer. A portion of the initial 100-fs laser pulses are directed to the reference arm containing an optical delay $\tau$. We monitor the transverse spatial intensity profile resulting from the interference of the synthesized STWP and plane-wave reference pulses via a CCD camera as we scan the optical delay $\tau$. When the two wave packets overlap in space and time, we observe spatially resolved interference fringes from which we reconstruct the local intensity profile at this axial plane $z$ and delay $\tau$. By scanning $\tau$ at a fixed axial plane, we reconstruct the time-resolved intensity profile $I(x,y,z;\tau)$. We plot in Fig.~\ref{fig:TimeResolvedIntensity}(b) this spatiotemporal intensity profile for the subluminal STWP [from Fig.~\ref{fig:TimeAverageData}(e)] at three axial planes, where the characteristic X-shaped intensity profile is clear. Similar results are shown in Fig.~\ref{fig:TimeResolvedIntensity}(c) for the superluminal STWP [from Fig.~\ref{fig:TimeAverageData}(f)].

The same measurement configuration in Fig.~\ref{fig:TimeResolvedIntensity}(a) can be used to assess the group velocity of the STWP. This is done by displacing the CCD camera along the propagation axis to a new position $z$. The spatially resolved fringes resulting from the interference of the STWP and the reference pulse are only observed when they overlap in space and time. However, the STWP travels at a group velocity $\widetilde{v}$, whereas the reference pulse travels at a group velocity $c$. Therefore, by displacing the CCD camera beyond the axial walk-off distance between the two wave packets, the spatial fringes disappear. High-visibility fringes are regained by inserting a delay $\tau$ in the reference arm to once again overlap the STWP and the reference pulse at the CCD camera in the new axial plane. From the displaced axial distance and the requisite added optical delay to retrieve the interference fringes, we can estimate the group velocity $\widetilde{v}$ of the STWP. We plot in Fig.~\ref{fig:TimeResolvedIntensity}(d) the measured delay against the axial displacement for the subluminal and superluminal STWPs, which lie below and above the luminal delays, respectively. We extend the measurements of $\widetilde{v}$ for other STWPs in Fig.~\ref{fig:TimeResolvedIntensity}(e) as we vary the parameter $B$ in the spectral re-organization stage. Both the subluminal and superluminal regimes can be accessed in this way.

\section{Discussion}

\subsection{New directions for investigations}

Previous work on STWPs in which the spatial spectrum was restricted to 1D in a Cartesian coordinate system was carried out with the universal AD-synthesizer described in Ref.~\cite{Hall24JOSAA,Romer25JOpt}, which can produce an arbitrary spatiotemporal spectrum $\widetilde{\psi}(k_{x};\Omega)$, but it cannot be generalized to a 2D spatial spectrum. The recent work on STWPs \cite{Yessenov22NC,Yessenov22OL,Yessenov25Meron} makes use of the spatiotemporal Fourier synthesizer described above. We briefly outline here extensions of this experimental strategy. 

\subsubsection{Spatiotemporal Fourier optics maintaining one-to-one correspondence between spatial and temporal frequencies}

We have focused here on propagation-invariant, group-velocity-tunable wave packets involving one-to-one correspondence between the spatial and temporal frequencies \cite{Bhaduri18OE,Kondakci19NC,Kondakci19ACSP,Bhaduri19OL,Hall25OE1km}. However, a host of other intriguing behaviors stem from engineering this one-to-one correspondence, including self-healing \cite{Kondakci18OL}, anomalous refraction \cite{Bhaduri19Optica,Bhaduri20NP,Motz21OL,Yessenov20JOSAA,Motz21JOSAA2,Yessenov21JOSAA3}, control over GVD in free space \cite{Yessenov21ACSPhot,Hall21PRAVwave} and in linear media \cite{Yessenov22OLDispersive}, dispersion cancellation \cite{Malaguti08OL,Malaguti09PRA,Dallaire09OE,Jedrkiewicz13OE,He22LPR,Hall23LPR,Hall23NatPhys,Hall24ACSP,Hall25PRA}, new classes of waveguide modes \cite{Shiri20NC,Guo21PRR,Shiri22Optica,Shiri22ACSP,Shiri23JOSAA}, and even omni-resonant fields in planar Fabry-P{\'e}rot cavities \cite{Shabahang17SR,Shiri20OL,Shiri20APLP,Villinger19unpubl,Hall25LRR,Shiri22OL}. Extending these realizations to azimuthally symmetric STWPs in which the one-to-one correspondence between spatial and temporal frequencies extends to both transverse dimensions is the next challenge for spatiotemporal Fourier optics.

Another intriguing direction for future study is exploring the connection between spatiotemporal Fourier optics and special relativity. The spatial spectrum for a monochromatic beam at $\omega_{\mathrm{o}}$ lies at the intersection of the hyper-cone $k_{x}^{2}+k_{y}^{2}+k_{z}^{2}=(\tfrac{\omega}{c})^{2}$ with the iso-frequency hyper-plane $\omega=\omega_{\mathrm{o}}$. It can be shown that the impact of a Lorentz boost is to convert this monochromatic beam into an STWP \cite{Belanger86JOSAA,Longhi04OE,Yessenov23PRA,Yessenov24PRA}. This perspective has indeed been instrumental in showing the equivalence \cite{Almeida25OL} between STWPs and the recently proposed `ideal flying focus' \cite{Ramsey23PRA}. Further study of the Lorentz boosts of conventional field configurations in two transverse dimensions is therefore expected to enrich spatiotemporal Fourier optics \cite{Saari04PRE,Bliokh12PRA,Bliokh2012PRL,Bliokh2013JOpt}.

\subsubsection{Discretized spatiotemporal spectra}

Discretizing the spatiotemporal spectrum leads to a host of phenomena related to the Talbot effect \cite{Talbot36PM}. Uniformly sampling the spatial spectrum of monochromatic light yields a spatially periodic field profile, whose profile is gradually lost with propagation, only to be reconstructed at equidistant axial planes separated by the spatial Talbot length $z_{\mathrm{T},x}=\tfrac{2L^{2}}{\lambda_{\mathrm{o}}}$, where $\lambda_{\mathrm{o}}$ is the wavelength and $L$ is the transverse period \cite{Berry96JMO}. A temporal Talbot effect exists for a periodic pulse train of period $T$ (corresponding to a uniformly sampled temporal spectrum) traveling in a dispersive medium with GVD coefficient $k_{2}$ \cite{Jannson81JOSA,Andrekson93OL,Mitschke98OPN}. Temporal dispersive spreading corrupts the field, only for the initial pulse train to be reconstructed axially in planes separated by the temporal Talbot length $z_{\mathrm{T},t}=\tfrac{T^{2}}{\pi|k_{2}|}$ \cite{Azana99AO}.

For spatiotemporally structured fields, the one-to-one correspondence between spatial and temporal frequencies, sampling the spatial spectrum entails simultaneously sampling the temporal spectrum, and vice versa. This gives rise to novel spatiotemporal Talbot effects, including a veiled space-time Talbot effect \cite{Yessenov20PRL}, a temporal Talbot effect in free space (in absence of a dispersive medium) \cite{Hall21OLTalbot}, and a true space-time Talbot effect \cite{Hall21APLSTTalbot} in which a space-time lattice (that is periodic in space and time) is lost under the combined impact of dispersion and diffraction, only for the initial lattice structure to be reconstructed at equidistant axial planes \cite{Hall21APLSTTalbot}. Extending the space-time Talbot effect to both transverse spatial dimensions provides a rich arena for self-imaging dynamics \cite{Zhang25OE}.

\subsubsection{Two-to-one spectral correspondence}

It is useful to consider fields endowed with a two-to-one correspondence; i.e., each spatial frequency is associated with two temporal frequencies. Rather than the open spectral surfaces in Figs.~\ref{fig:PulsedBessel}, \ref{fig:XwaveCones}, \ref{fig:SubluminalCones}, and \ref{fig:SuperluminalCones} realized with a one-to-one correspondence, \textit{closed} spectral surfaces necessitate a two-to-one correspondence. This allows studying topological spin textures of the polarization vector on closed \cite{Yessenov25Meron} or open \cite{Guo21Light} spectral surfaces.

Another line of investigation that benefits from a two-to-one correspondence is producing so-called linear O-waves in the anomalous GVD regime \cite{Malaguti08OL}. As shown earlier, X-waves and STWPs have X-shaped spatiotemporal profiles in any meridional plane (see Refs.~\cite{Kondakci18PRL,Wong21OE} for exceptions in presence of spectral phase modulation). However, it has been theorized that an O-shaped STWP can exist in presence of anomalous GVD \cite{Malaguti08OL}, whereupon the spatiotemporal spectrum takes the form of a circle involving a two-to-one correspondence, yielding an STWP whose spatiotemporal profile in physical space is circularly symmetric in space and time. This has been verified with spatial spectra restricted to one dimension \cite{Hall24ACSP}. Moreover, these O-waves are an optical realization of the long-sought de~Broglie-Mackinnon wave packets \cite{Mackinnon78FP,Hall23NatPhys,Diouf23NP}, which are propagation-invariant solutions of the Klein-Gordon equation. Finally, tuning the group velocity of an STWP in a medium in its anomalous-GVD regime can display an abrupt transition from an X-shaped to an O-shaped spatiotemporal profile \cite{Hall24ACSP}.

Extending this work to 2D spatial spectra will produce O-waves that are spherically symmetric in space and time, which may be useful for coupling to optical fibers (in the anomalous-GVD regime in the telecommunications window), and producing the localized wave packets with spherical-harmonic symmetries proposed in Ref.~\cite{Mills12PRA}. Realizing a two-to-one correspondence between temporal and spatial frequencies requires modifying the 1D spectral transformation stage in the spatiotemporal Fourier synthesizer described above (Fig.~\ref{fig:OverallSetup} and Fig.~\ref{fig:1DTransform}), a particular realization of which was reported recently in Ref.~\cite{Yessenov25Meron}.

\subsubsection{Finite-bandwidth spatiotemporal spectral association}

Novel phenomena emerge once the strict one-to-one or even two-to-one correspondence between spatial and temporal frequencies is abandoned. By associating a prescribed complex-valued spatial spectrum of narrow spatial bandwidth to each temporal frequency, controlled deviation from propagation invariance can be realized. Examples include axially accelerating pulses \cite{Clerici08OE,Lukner09OE,Yessenov2020PRLaccel,Li20SR,Li20CP,Li21CP,Hall22OLAccel}, varying wavelength of the on-axis field along the propagation axis (axial spectral encoding) \cite{Motz20arxiv}, and potentially even curved-beam propagation \cite{Liang23OL}.

Although some of these phenomena have been observed with 1D spatial Fourier spectra, extending these realizations to the more general 2D Fourier spatial spectra considered here involves a technical challenge. Specifically, the 1D spectral transformation is a conformal mapping that transmits the field at any position $x_{1}$ in the input plane to a unique position $x_{2}$ in the output plane. To realize field configurations in which each position $x_{1}$ (associated with a single $\omega$) is mapped to a prescribed width $\Delta x$ at a target position $x_{2}$ in the output plane is a task that has not yet been achieved, and remains a future challenge.

\subsubsection{Full spatiotemporal Fourier optics}

The most general non-separable spatiotemporal spectrum $\widetilde{\psi}(k_{r},\Omega)$ for an azimuthally symmetric pulsed field conforms to none of the above cases. Synthesizing such a spatiotemporal spectrum requires the ability to encode an arbitrary temporal spectrum at each radial location in the Fourier plane; or vice versa, associating an arbitrary radial distribution to each temporal frequency. Constructing an optical system capable of producing such arbitrary spatiotemporal spectra can be utilized to explore the swathe of spatiotemporally structured optical fields that are currently being investigated \cite{Yessenov22AOP}. 

Moreover, the spatiotemporal spectrum $\widetilde{\psi}(k_{r},\chi,\Omega)$ has been contracted here to azimuthally symmetric fields $\widetilde{\psi}(k_{r},\chi,\Omega)\rightarrow\widetilde{\psi}(k_{r},\Omega)$. Retaining full control over the spatiotemporal spectrum requires the additional ability to manipulate the spectrum along its azimuthal dimension $\chi$. This can be achieved in the 1D spectral transformation stage by exploiting the unused dimension $y$. The phase plates used in this stage are modulated only along $x$ (Fig.~\ref{fig:1DTransform}). The Cartesian dimension $y$ is subsequently rolled by the 2D coordinate transformation into the angular coordinate. Adding a linear phase along $y$ (spanning integer multiples of $2\pi$) does not impact the 1D spectral transformation, but will endow the spectrum with OAM \cite{Yessenov22NC}, resulting in OAM-carrying, propagation-invariant STWPs. The consequences of extending this OAM strategy to arbitrary spatiotemporal spectra in $k_{r}$ and $\Omega$ remains an open question.

\begin{figure}[t!]
    \centering
    \includegraphics[width=6.6cm]{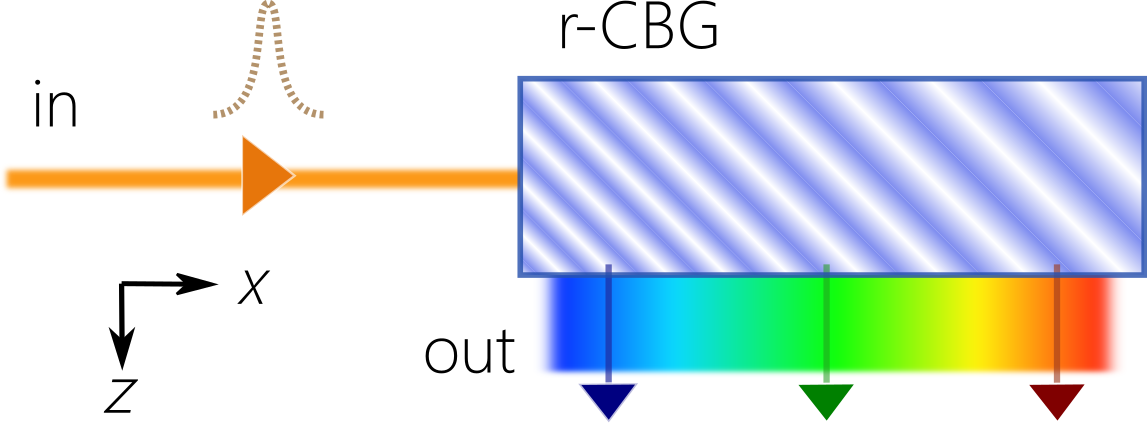}
    \caption{Schematic of a rotated-CBG (r-CBG). A collimated, broadband, spatially narrow field is incident on one facet of the device, and the spatially resolved spectrum emerges from a device facet that is perpendicular to the input facet.}
    \label{fig:rCBG}
\end{figure}

\subsection{Advances in spectral analysis}

As mentioned earlier, CBGs can alleviate some of the difficulties associated with conventional gratings \cite{Mhibik23OL,Mhibik23OL2}. In one realization, the Bragg structure is written in the device volume tilted by an angle $45^{\circ}$ with respect to the facets [Fig.~\ref{fig:rCBG}]. This rotated CBG (r-CBG) allows for reducing the grating volume for the same resolved bandwidth, and it allows for the initial pulse to impinge normally on the input facet, and for the spatially resolved spectrum to exit normally from a different facet \cite{Mhibik23OL}. Such a device can dramatically simplify the setup in Fig.~\ref{fig:OverallSystem}, and it has already shown utility in reducing the volume of the setup for synthesizing STWPs that are localized along one transverse spatial dimension (ST light-sheets) \cite{Yessenov23OL}.

\subsection{Spatiotemporal Fourier synthesis with OAM modes}

The above-described spatiotemporal Fourier synthesis system is capable of producing almost arbitrary azimuthally symmetric, radial distributions for the spectrum in the Fourier plane, which is limited primarily by the system aperture. Difficulties can be faced however if the spectrum is discrete, as in the case of using a laser comb as the source. An intriguing case was studied theoretically in Ref.~\cite{Pariente15OL}, where each temporal frequency is associated with a Laguerre-Gaussian mode, resulting in a wave packet with helical intensity and phase profiles, a field structure that was thus termed a spatiotemporal light spring. In contrast to the space-time Talbot effect in which each temporal frequency is associated with a spatial frequency from the continuum of plane-wave modes (each of which is propagation invariant), this alternative approach selects the spatial modes from a discrete set (Laguerre-Gaussian modes, which are not themselves diffraction-free). Consequently, spatiotemporal light springs are not propagation-invariant wave packets, but can still display useful features such as a tunable group velocity. Complex propagation dynamics follow from associating each spectral line with a precisely engineered superposition of OAM modes \cite{Zhao20NC}.

\begin{figure}[t!]
    \centering
    \includegraphics[width=8.6cm]{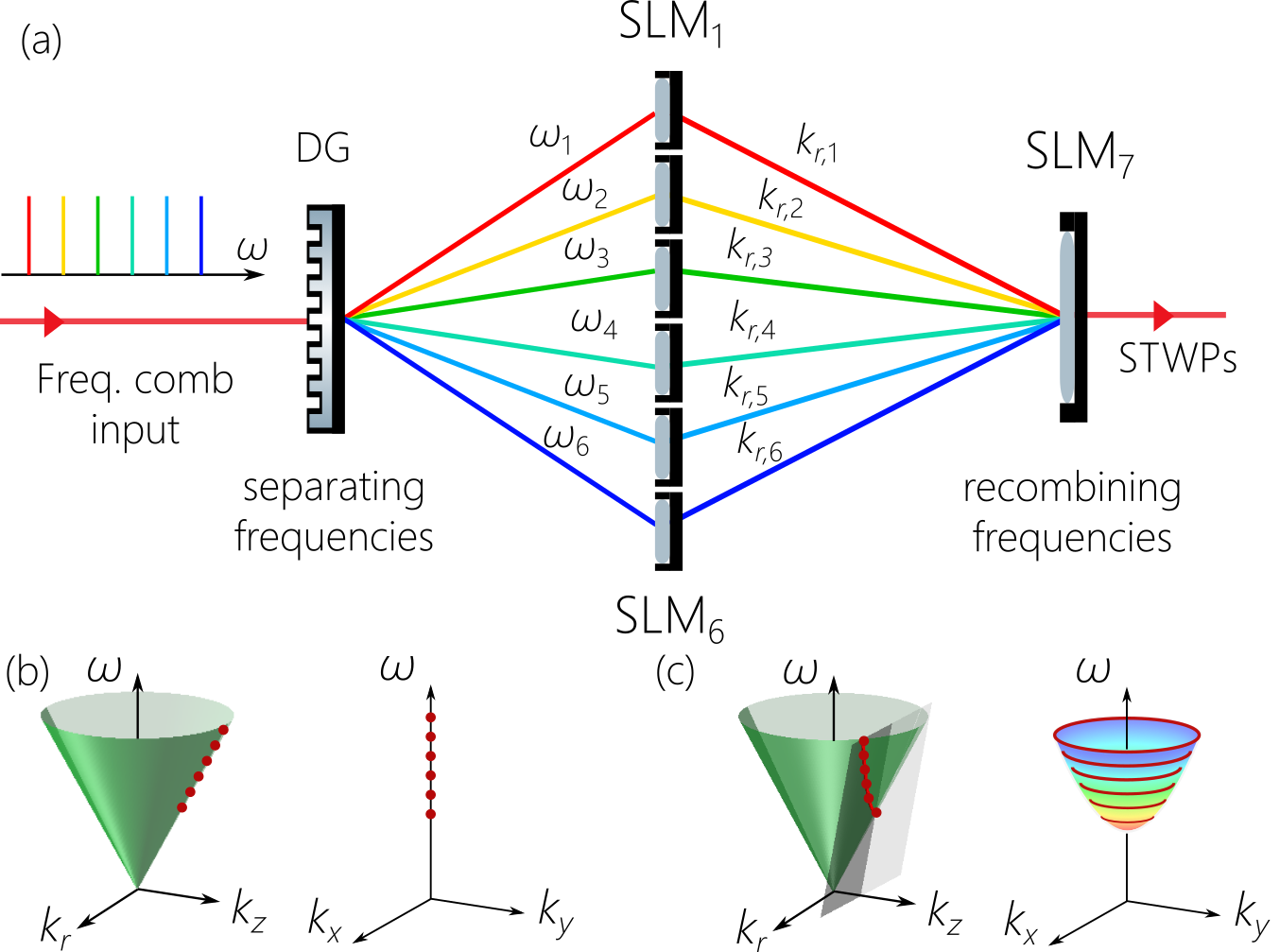}
    \caption{(a) A spatiotemporal spectral modulator based on an array of SLMs. A diffraction grating (DG) spatially resolves the spectrum of a frequency comb, each spectral line is incident on a different SLM, or a different section of an SLM, where it is spatially modulated. (b) Spectral support of a plane-wave frequency comb on the surface of the light-cone. (c) Spectral support of the frequency comb after introducing the spatiotemporal structure corresponding to a superluminal STWP.}
    \label{fig:WillnerExperiment}
\end{figure}

This approach, in which each spectral line in a comb is associated with an OAM mode, has been developed by A. Willner \textit{et al.} using the strategy illustrated in Fig.~\ref{fig:WillnerExperiment}. The discrete spectral lines in the comb are separated spatially by a grating, and each temporal frequency is directed to a portion of an SLM on which a prescribed spatial distribution is imparted [Fig.~\ref{fig:WillnerExperiment}(a)], whether Laguerre-Gauss or Bessel modes of different OAM order, or superpositions of such modes \cite{Pang22OE,Pang22OL,Zou22OL,Minoofar22OE,Zou23OL,Su24OE}. The spectral support of a plane-wave frequency comb on the light-cone surface is discrete and lies on the light-line [Fig.~\ref{fig:WillnerExperiment}(b)], and after the spatiotemporal spectral modulation to produce a superluminal STWP [Fig.~\ref{fig:WillnerExperiment}(a)], the spectral support is a discretized version [Fig.~\ref{fig:WillnerExperiment}(c)] of the superluminal STWP in Fig.~\ref{fig:SuperluminalCones}.

\subsection{Further developments in spatiotemporal Fourier optics}

We have focused here on conical-AD, which is a special case of spatiotemporal Fourier optics in which each temporal frequency $\omega$ is associated with a single spatial frequency $k_{r}$. This is not the most general case; indeed, many recently developed spatiotemporally structured fields do \textit{not} belong to this class. Examples of these other fields include STOVs \cite{Jhajj16PRX,Hancock19Optica,Chong20NP,Hancock21Optica} that are prepared with a spatiotemporal Fourier synthesizer exploiting an SLM in which one dimension was reserved for the wavelength and another for a 1D spatial spectrum. Consequently, the STOV field is modulated along one transverse spatial dimension and remains uniform along the other, similar to early work on STWPs. Toroidal pulses \cite{Wan22NP,Zdagkas22NP} are produced from an STOV after rolled into a toroid using the log-polar coordinate transformation in Fig.~\ref{fig:2DTransformConcept} and Fig.~\ref{fig:2DPhasePlate}. The flying focus \cite{SaintMarie17Optica,Froula18NP,Jolly20OE}, each of is produced by chromatic focusing of a stretched (chirped) optical plane-wave pulse, resulting in a focal spot whose on-axis wavelength changes with propagation and travels at a tunable group velocity \cite{Froula18NP}. Exploiting an axiparabola-echelon pairing has been reported to produce an `achromatic' flying focus, but with a reduced tunability of the group velocity \cite{Pigeon24OE}.

Other recent approaches include using a metasurface to replace the SLMs in the configuration depicted in Fig.~\ref{fig:WillnerExperiment}(a) after spatially resolving the temporal spectrum, to endow each wavelength (in case of a frequency-comb source) or portion of the spectrum (in case of a source with a continuous spectrum) with an arbitrary complex vector field \cite{Chen22SciAdv}. In another approach, a combination of circular grating (to spatially resolve the spectrum radially) and a metasurface (to spatially modulated the spectrally resolved wavefront) has been used to produce STWPs in which the spatiotemporal field topology is modified \cite{Piccardo23NP}. Another experimental strategy relies on an MPLC to introduce a wavelength-dependent spatial profile \cite{Mounaix20NC,CruzDelgado22NP}. A host of other approaches are being pursued to introduce controllable space-time coupling into the field \cite{Chen22NC,Lin24NC,Huo24NC,Liu24NC,Cao24NC,Huang24SciAdv,Chen25NC}, but which are being developed piecemeal. It remains an open challenge whether a general spatiotemporal Fourier synthesizer can be devised to encompass the entire space of spatiotemporally structured light.

\section{Conclusions}

In conclusion, we have presented a formulation for the nascent field of spatiotemporal Fourier optics. In contrast to conventional \textit{spatial} Fourier optics used in beam shaping and conventional \textit{temporal} Fourier optics used in ultrafast pulse-shaping, the emerging field of \textit{spatiotemporal} Fourier optics requires joint modulation of the spatial and temporal spectra to produce spatiotemporal spectra that are \textit{not} separable along the spatial and temporal dimensions. We have addressed here a subspace of the full space of spatiotemporally structured fields, namely azimuthally symmetric fields in which each radial spatial frequency is associated with a single temporal frequency (conical-AD). This subclass is endowed with a diffraction-free time-averaged spatial intensity profile, a subset of which are STWPs that are propagation invariant: the wave packet travels rigidly in free space without diffraction or dispersion at a tunable group velocity. We have presented a theoretical formulation of spatiotemporal Fourier optics for this class of fields that emphasizes the utility of propagation angles rather than spatial frequencies to avoid mixing the spatial and temporal degrees-of-freedom. 

We have described an optical system for spatiotemporal Fourier synthesis capable of producing almost arbitrary spectral configurations in this subclass of fields, along with spectral, spatial, and spatiotemporal characterization schemes. Spatiotemporal Fourier optics is still early in its development, but the new advances currently being reported bodes well for future growth in fundamental concepts and potential applications \cite{Piccardo22JO,Shen23JO}.

\subsection*{Funding}
U.S. Office of Naval Research (ONR) N00014-20-1-2789.

\subsection*{Disclosures}
The authors declare no conflicts of interest.

\subsection*{Data availability}
Data underlying the results presented in this paper are not publicly available at this time but may be obtained from the authors upon reasonable request.

\bibliography{diffraction}


\end{document}